\documentclass[11pt]{article}

\usepackage[utf8]{inputenc}
\usepackage{amsmath,amsthm,amsfonts,amssymb}
\usepackage{graphicx}
\usepackage[authoryear,sort]{natbib}

\RequirePackage{xcolor}
\definecolor{goldenbrown}{rgb}{0.6, 0.4, 0.08}
\RequirePackage[colorlinks,linkcolor=goldenbrown,citecolor=goldenbrown,urlcolor=goldenbrown]{hyperref}

\usepackage{fancyhdr}
\usepackage{caption}
\usepackage[right=1.2in,left=1.2in,top=.8in,bottom=.8in]{geometry}
\usepackage{algorithm}

\usepackage[explicit]{titlesec} 
\titleformat{\section}{\normalfont\large\bf}{\thesection}{1em}{#1.}
\titleformat{\subsection}[runin]
  {\normalfont\normalsize\bf}{\thesubsection}{1em}{#1.}
\titleformat{\subsubsection}[runin]
  {\normalfont\normalsize\it}{\thesubsubsection}{1em}{#1.}

\usepackage{lmodern}

\date{}

\newcommand{\E}{\mathbb{E}}

\newcommand{\cov}{{\rm Cov}}
\newcommand{\pto}{\stackrel{P}{\to}}
\newcommand{\dto}{\stackrel{D}{\to}}
\newcommand{\iid}{\stackrel{{\rm i.i.d.}}{\sim}}

\newcommand{\T}{{\mathrm T}}
\numberwithin{equation}{section}
\theoremstyle{plain}
\newtheorem{theorem}{Theorem}[section]
\newtheorem{definition}{Definition}[section]
\newtheorem{corollary}{Corollary}[section]
\newtheorem{lemma}{Lemma}[section]

\usepackage{float}

\newlength\FHoffset
\fancyheadoffset{\FHoffset}
\title{\vspace{-10mm}{\bf \large LINKED FACTOR ANALYSIS}}
\author{\sc Giuseppe Vinci\\{\footnotesize Department of Applied and Computational Mathematics and Statistics}\vspace{-2mm}\\  {\footnotesize  University of Notre Dame, Notre Dame, Indiana, USA}}

\begin{document}
\maketitle
 \vspace{-7mm}
\begin{abstract}
\footnotesize Factor models are widely applied to the analysis of multivariate data across disparate fields of research. However, modern scientific data are often incomplete, and estimating a factor model from partially observed data can be very challenging. In this work, we show that if the data are structurally incomplete, the factor model likelihood function can be decomposed into a product of likelihood functions for multiple factor models relative to different observed data subsets. If these factor models are linked together by common parameters, we can obtain complete maximum likelihood estimates of the full factor model parameters. We call this modeling framework Linked Factor Analysis (LINFA). LINFA can be used for covariance matrix completion, dependence estimation, dimension reduction, and data completion. We compute the maximum likelihood estimator through an efficient Expectation-Maximization algorithm, accelerated by a novel Group Vertex Tessellation algorithm. We establish the conditions for the consistency and asymptotic normality of the estimator. We design confidence regions, hypothesis tests, bootstrap algorithms, and methods for selecting the number of factors. Finally, we illustrate the application of LINFA in an extensive simulation study and in the analysis of neuroscience data.
\end{abstract}

\noindent%

\section{Introduction}\vspace{-3mm}
Factor models have been widely applied in the analysis of multivariate data in disparate fields of research, including neuroscience \citep{bong2020latent,freeman1987relation,vinci2018adjusted}, psychology \citep{russell2002search}, psychometrics \citep{barratt1965factor}, econometrics \citep{onatski2009testing,ahn2013eigenvalue}, financial econometrics \citep{chamberlain1982arbitrage,fama1992cross,campbell1996understanding,stock2002forecasting,onatski2012asymptotics},  macroeconomics \citep{stock2002macroeconomic,boivin2006more}, and astrophysics \citep{patat1994light,kurtz2000nasa}. In a classical Gaussian exploratory factor analysis model, we observe $n$ independent and identically distributed (i.i.d.)~$d$-dimensional random vectors $X_1, \ldots, X_n$, generated according to the equation
\begin{equation}\label{eq:simpleFA}
X_r=\mu+\Lambda Z_r  + \varepsilon_r,
\end{equation}
where $ \mu\in\mathbb{R}^d$ is the mean vector, $\Lambda\in\mathbb{R}^{d\times q}$ is the loading matrix, $Z_1$, \ldots, $Z_n\iid N(0,I_q)$ are $q$-dimensional latent factors with an identity covariance matrix, and $\varepsilon_1$, $\ldots$, $\varepsilon_n\iid N(0,\Psi)$ are $d$-dimensional error components with a diagonal covariance matrix $\Psi$. Indeed, $X_1, \ldots,X_n\iid N(\mu,\Sigma)$, where
\begin{equation}\label{eq:cov}
\Sigma=\Lambda\Lambda^\T+\Psi
\end{equation}
is a $d\times d$ covariance matrix, and  $\Lambda\Lambda^\T$ is a $d\times d$ matrix of rank no larger than $ q $. The parameters $\mu$, $\Lambda$, and $\Psi$ of the factor model can be estimated in various ways, including  
maximum likelihood estimation  \citep{bai2012statistical,doz2012quasi}, 
penalized maximum likelihood estimation  \citep{carvalho2008high,tsai2010constrained,kneip2011factor,caner2014selecting}, and Bayesian methods \citep{bhattacharya2011sparse,knowles2011nonparametric,rovckova2016fast,srivastava2017expandable}.

Factor models are very attractive from the estimation point of view because the number of parameters involved can be maintained substantially low even for very large $d$. Indeed, the total number of parameters in the components $\Lambda$ and $\Psi$ that characterize the covariance matrix $\Sigma$ in Equation~\eqref{eq:cov} is $d(q+1)$, which can be much smaller than the full dimensionality $d(d+1)/2$ of an unconstrained $d\times d$ covariance matrix if $q< (d-1)/2$. This property is especially useful in the high-dimensional setting since, for a fixed number of factors $q$, the total number of parameters in a factor model increases only linearly with $d$, while in the unconstrained case it increases quadratically. However, Equation~\eqref{eq:cov} is notoriously affected by rotational invariance: for any $q\times q$ orthogonal matrix $P$, we have that $\tilde\Lambda=\Lambda P$ yields the same covariance matrix $\Sigma=\tilde\Lambda\tilde\Lambda^\T+\Psi$ simply because $\Lambda\Lambda^\T=\tilde\Lambda\tilde \Lambda^\T$. With no appropriate constraints, any estimation algorithm would yield one of infinitely many possible estimates of $\Lambda$ \citep{anderson1956statistical,lawley1962factor, bai2012statistical}. Nevertheless, the covariance matrix $\Sigma$ in Equation~\eqref{eq:cov} is unaffected by the rotational invariance problem. 

In this paper, we investigate factor model estimation under \textit{structural} or \textit{deterministic missingness}. This setting is characterized by incompletely observed data vectors $X_1, \ldots, X_n$, where certain variable pairs may never be observed jointly, thereby precluding the computation of a full sample covariance matrix. For example, in neuroscience, it is often impossible to record the activity of an entire population of neurons simultaneously with a reasonable temporal resolution. Instead, smaller subsets of neurons are usually recorded simultaneously over independent experimental sessions using fine temporal resolution. These settings, however, yield $K$ data sets about the activity of $K$ different subsets of neurons, and several pairs of neurons may never be recorded jointly. Structural or deterministic missingness profoundly differs from typical missing-data problems \citep{candes2009exact, candes2010matrix, loh2011high, kolar2012estimating, stadler2012missing, soudry2015efficient}, such as missingness at random and its variants, where every variable pair is observed at least a few times with high-probability. The type of structural missingness we deal with has been investigated outside of factor modeling, such as in the context of sparse precision matrix estimation \citep{vinci2019graph}, and recovery of exact low rank positive semidefinite matrices \citep{bishop2014deterministic}.

We tackle the problem of estimating a factor model, in particular the covariance matrix $\Sigma$ in Equation~\eqref{eq:cov}, from structurally incomplete data as follows. 
In Section~\ref{sec:linfa}, we show that the likelihood function of the observed data can be factorized into the product of the likelihood functions of multiple partial factor models that are linked together by shared portions of $\Lambda$ and $\Psi$. 
We call this modeling framework \textit{Linked Factor Analysis} (LINFA) and demonstrate how it can be used for covariance matrix completion, dimension reduction, data completion, and dependence structure recovery in the case of structural data missingness. 

In Section~\ref{sec:mle}, we study the maximum likelihood estimation (MLE) of the parameters $(\Lambda,\Psi)$ of LINFA. In Section~\ref{sec:linkage}, we establish the \textit{linkage condition} for the uniqueness of the MLE of $\Sigma$. In Section~\ref{sec:EM}, we propose an efficient \textit{Expectation-Maximization} (EM) algorithm for the  computation of the MLE. 
This algorithm can be applied to data with arbitrary missingness patterns and involves updating equations that are all in closed form. Moreover, the M step of the algorithm is accelerated by a novel \textit{Group Vertex Tessellation} (GVT) algorithm, which identifies a minimal partition of the vertex set to implement efficient maximization steps. The GVT algorithm is very fast and needs to be run only once before the EM algorithm is executed. In Section~\ref{sec:theory}, we establish the conditions for the consistency and asymptotic normality of the LINFA MLE and propose confidence regions, hypothesis tests, and bootstrap methods. 
Finally, in Section~\ref{sec:modelselection}, we consider three approaches for the empirical selection of the number of factors in LINFA: cross-validation, Akaike information criterion, and Bayesian information criterion. 

In Section~\ref{sec:simulations}, we demonstrate via simulation that our proposed methods for uncertainty quantification and model selection for LINFA are adequate. We further demonstrate that our proposed EM algorithm with GVT acceleration is computationally efficient and that LINFA can be effectively used for covariance matrix completion, dimension reduction, data completion, and dependence structure recovery. In Section~\ref{sec:data}, we apply the methods to the analysis of neuronal calcium imaging data collected from a mouse visual cortex. Finally, in Section~\ref{sec:discussions}, we discuss our results and future research directions.

\section{The Linked Factor Analysis Model}\label{sec:linfa}\vspace{-3mm}

\subsection{Framework}\label{sec:linfaframework}
Assume the generative factor model in Equation~\eqref{eq:simpleFA}, and suppose we observe $K$ independent data matrices $\mathbf{X}^{(1)}, \ldots, \mathbf{X}^{(K)}$, where $\mathbf{X}^{(k)}\in\mathbb{R}^{n_k\times |V_k|}$ contains $n_k$ independent and identically distributed samples $X_1^{(k)}, \ldots,X_{n_k}^{(k)}$ about variables in the set $V_k$, where $V_1,\ldots,V_K\subset V=\{1,\ldots,d\}$ are distinct nonempty subsets and $\cup_{k=1}^K V_k=V=\{1,\ldots,d\}$. 
Thus, for $r=1,\ldots,n_k, ~k=1,\ldots,K$, 
\begin{equation}
X_r^{(k)} = \mu^{(k)}+\Lambda^{(k)}Z_r^{(k)}+\varepsilon_r^{(k)},
\end{equation}
where $\mu^{(k)}=(\mu_j)_{j\in V_k}$,  $\Lambda^{(k)}=\Lambda_{V_k}$, $\Lambda_U:=[\Lambda_{ij}]_{i\in U,j\in\{1,\ldots,q\}}$,  $\varepsilon_1^{(k)},\ldots,\varepsilon_{n_k}^{(k)}\iid N(0,\Psi^{(k)})$, and $\Psi^{(k)}=[\Psi_{ij}]_{i,j\in V_k}$. Consequently, $X_r^{(k)}\sim N(\mu^{(k)}, \Sigma^{(k)})$, where 
\begin{equation}\label{eq:covk}
\Sigma^{(k)}=\Lambda^{(k)}\Lambda^{(k)\T}+\Psi^{(k)}
\end{equation}
For simplicity, we shall assume $\mu=0$ (in practice, we can center the data by subtracting the vector of sample means computed with the available data). We further assume that the observational pattern is structural and independent of any random variable in the system. Therefore, the likelihood function of the observed data $\mathbf{X}^{(1)}, \ldots, \mathbf{X}^{(K)}$ takes the form
\begin{equation}\label{eq:linfalikelihood}
L(\Lambda,\Psi; \mathbf{X}^{(1)},\ldots, \mathbf{X}^{(K)})  ~=~ \prod_{k=1}^K L_k(\Lambda^{(k)},\Psi^{(k)};\mathbf{X}^{(k)}),
\end{equation}
where $L_k(\Lambda^{(k)},\Psi^{(k)};\mathbf{X}^{(k)})$ is the Gaussian likelihood function for the data set $\mathbf{X}^{(k)}$. In Section~\ref{sec:mle}, we establish the \textit{linkage condition}, which prescribes the minimum overlap among the sets $V_1,\ldots,V_K$ to enable the recovery of full estimates for $\Lambda$, $\Psi$, and $\Sigma$ through LINFA likelihood maximization.

\begin{figure*}[t!]
\centering
\includegraphics[width=1\textwidth]{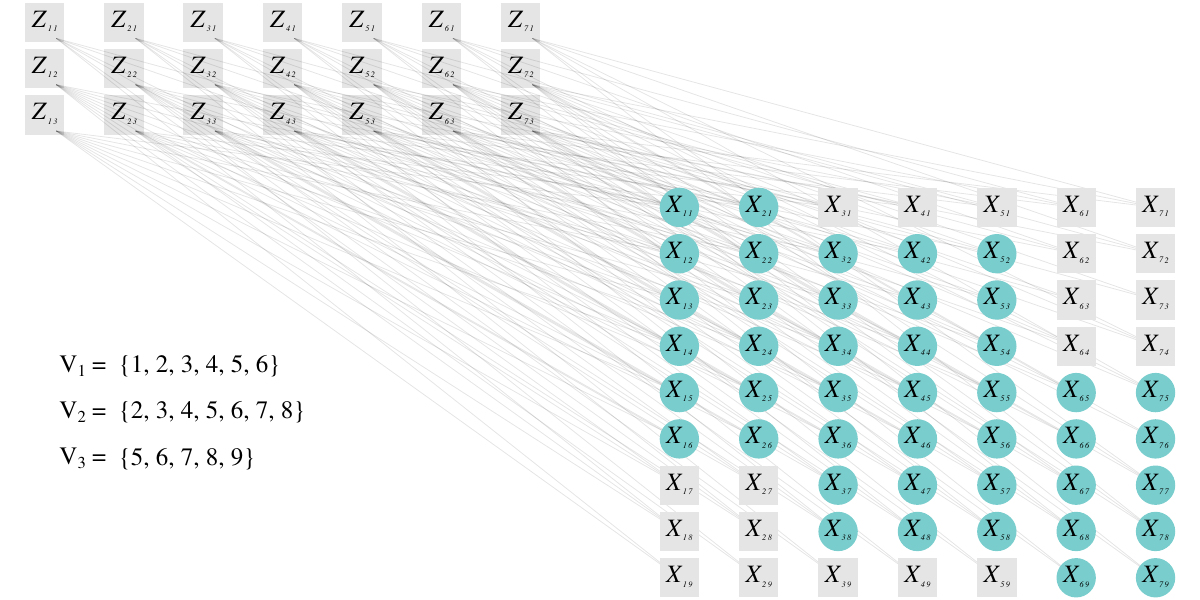}
\caption{\textbf{Structural missingness in factor analysis:} $9$-dim random vectors ($X$'s) depend on $3$ latent factors ($Z$'s). Only the variables denoted by blue circles are observed.}\label{fig:intro}
\end{figure*}

Figure~\ref{fig:intro} shows an example where $q=3$, $d=9$, $V_1=\{1,\ldots,6\}$, $V_2=\{3,\ldots,8\}$, and $V_3=\{5,\ldots,9\}$. It can be seen that the sub-matrices $\Lambda^{(1)}$ and $\Lambda^{(2)}$ share the entries $[\Lambda_{ij}]_{i\in\{3,\ldots,6\},j\in \{1,\ldots,q\}}$, and $\Psi^{(1)}$ and $\Psi^{(2)}$ share the entries $\{\Psi_{ii}\}_{i\in\{3,\ldots,6\}}$. Similarly, $\Lambda^{(2)}$ and $\Lambda^{(3)}$ share the entries $[\Lambda_{ij}]_{i\in\{5,\ldots,8\},j\in \{1,\ldots,q\}}$, and $\Psi^{(2)}$ and $\Psi^{(3)}$ share the entries $\{\Psi_{ii}\}_{i\in\{5,\ldots,8\}}$. Note that the variable pairs $ (1,7)$, $(1,8)$, $(1,9)$, $(2,9)$, $(3,9)$, $(4,9)$ have no joint observations, which would prevent us from computing a complete sample covariance matrix. However, the sets $V_1,V_2,V_3$ overlap sufficiently (in the sense described in Section~\ref{sec:linkage}) to allow for the complete maximum likelihood estimation of $\Lambda$ and $\Psi$, and thereby $\Sigma$.

\subsection{Applications}
LINFA is a unified framework for \textit{covariance matrix completion}, \textit{dimensionality reduction}, \textit{data completion}, and \textit{dependence structure recovery} in the case of structural missingness. Given estimates $(\hat\Lambda,\hat\Psi)$, the covariance matrix estimate is 
\begin{equation}
\hat\Sigma = \hat\Lambda\hat\Lambda^\T+\hat\Psi
\end{equation}
Moreover, dimensionality reduction can be performed by predicting the latent factors as
\begin{equation}\label{eq:factorpred}
\hat Z_r^{(k)} ~=~ \hat\Lambda^{(k)\T}\hat\Sigma^{(k)-1}X_r^{(k)},
\end{equation}
where $\hat\Lambda^{(k)}$ is the estimate of $\Lambda^{(k)}$,  $\hat\Sigma^{(k)}=\hat\Sigma_{V_kV_k}$, and $X_r\in\mathbb{R}^{|V_k|}$ is the $r$-th sample in $\mathbf{X}^{(k)}$. Furthermore, data completion can be implemented by using the relevant predicted entries in
\begin{equation}\label{eq:datacompl}
\hat X_r^{(k)}=\hat\Lambda \hat Z_r^{(k)}, 
\end{equation}
Finally, conditional dependence can be studied in two ways: (a) unconditionally on the factors and (b) conditionally on the factors. In case (a), we study the partial correlations between variable pairs $(X_{ri},X_{rj})$, 
\begin{eqnarray}\label{eq:rho}
\rho_{ij} &:=& {\rm Cor}(X_{ri},X_{rj}\mid \{X_{rh}\}_{h\notin\{i,j\}})\\
&=& -\frac{\Theta_{ij}}{\sqrt{\Theta_{ii}\Theta_{jj}}},
\end{eqnarray}
for $1\le i<j\le d$, where $\Theta=\Sigma^{-1}$ can be computed via Woodbury Matrix Identity 
\begin{equation}\label{eq:invcovwood}
\Theta~=~\Psi^{-1}-\Psi^{-1}\Lambda(I+\Lambda^\T\Psi^{-1}\Lambda)^{-1}\Lambda^\T\Psi^{-1}
\end{equation}
Then, a partial correlation graph will contain $d$ nodes representing the variables $X_{r1}$, $\ldots$, $X_{rd}$, and an edge connects nodes $(i,j)$ with strength proportional to the magnitude of $\rho_{ij}\in(-1,1)$. In case (b), we compute the conditional correlations
\begin{eqnarray}\label{eq:gamma}
    \gamma_{ij} &=& {\rm Cor}(X_{ri},Z_{rj}\mid \{Z_{rh}\}_{h\neq j}) \\
    &=&\frac{\Lambda_{ij}}{\sqrt{\Lambda_{ij}^2+\Psi_{ii}}},
\end{eqnarray}
for $ 1\le i\le d, 1\le j\le q$. Then, we can  construct a factor bipartite graph with $d$ nodes representing the variables $X_{r1},\ldots,X_{rd}$, $q$ nodes representing the latent factors $Z_{r1},\ldots,Z_{rd}$, and with an edge connecting a variable node $X_{ri}$ to a latent node $Z_{rj}$ with strength proportional to the magnitude of $\gamma_{ij}\in (-1,1)$; no edge connects any two variables $X_{ri},X_{rj}$ or any two latent factors $Z_{ri},Z_{rj}$.  We explore all these applications of LINFA in Section~\ref{sec:simulations}.

\section{Maximum Likelihood Estimation}\label{sec:mle}\vspace{-3mm}

In this section, we propose an expectation-maximization (EM) algorithm to compute the maximum likelihood estimator
\begin{equation}\label{eq:MLE}
   (\hat\Lambda,\hat\Psi)~=~ \underset{\Lambda\in\mathbb{R}^{d\times q},\Psi\in \mathcal{D}^{d\times d}_{++}}{\arg\max} L(\Lambda,\Psi; \mathbf{X}^{(1)},\ldots, \mathbf{X}^{(K)}),
\end{equation}
where $L(\Lambda,\Psi; \mathbf{X}^{(1)},\ldots, \mathbf{X}^{(K)}) $ is the likelihood function defined in Equation~\eqref{eq:linfalikelihood}, and $\mathcal{D}^{d\times d}_{++}$ is the set of $d\times d$ positive diagonal matrices. Our EM algorithm can be applied to data with an arbitrary missingness pattern as long as some conditions are met (Section~\ref{sec:linkage}), and it involves updating equations all in closed form. Moreover, our EM algorithm is accelerated by the novel Group Vertex Tessellation (GVT) algorithm, which identifies a minimal partition of the vertex set to implement efficient maximization steps. The GVT algorithm needs to be run only once before the full EM algorithm is executed. 

In Section~\ref{sec:linkage}, we establish the linkage condition for the uniqueness of the LINFA MLE. In Section~\ref{sec:EM}, we present our EM algorithm. In Section~\ref{sec:theory}, we establish the consistency and asymptotic normality of the LINFA MLE and propose confidence regions, hypothesis tests, and bootstrap methods.
Finally, in Section~\ref{sec:modelselection}, we propose three approaches for the empirical selection of the number of factors.

\subsection{Linkage condition}\label{sec:linkage} 
LINFA, like other factor models, faces the challenge of rotational invariance in its loading matrix $\Lambda$. We mitigate this by imposing the canonical rotation constraint 
\begin{equation}\label{eq:canon}
\Lambda^\T\Psi^{-1}\Lambda = D\in\mathcal{D}_{++}^{d\times d},~  ~0<D_{qq}<\ldots<D_{11}<\infty,
\end{equation}
which ensures the identifiability of $\Lambda$ up to column sign flips \citep{rao1955estimation,jennrich1974simplified}. However, rotational invariance does not impede the full identification of $\Sigma$. A more significant concern for LINFA arises when independent, arbitrary rotations are possible across different subsets of $\Lambda$. Indeed, the LINFA likelihood function in Equation~\eqref{eq:linfalikelihood} depends on $(\Lambda,\Psi)$ only through the components $\Sigma^{(1)},\ldots,\Sigma^{(K)}$, where $\Sigma^{(k)}=\Lambda^{(k)}\Lambda^{(k)\T}+\Psi^{(k)}$. Thus, if $(\tilde\Lambda,\tilde\Psi)$ is an MLE, then any solution $(\breve\Lambda,\breve\Psi)$ that yields 
\begin{equation}\label{eq:mleequiv1}
\breve\Sigma^{(k)}=\tilde\Sigma^{(k)},~\forall k=1,\ldots,K
\end{equation}
is also an MLE. However, while $\breve\Sigma_O=\tilde\Sigma_O$, where $O=\cup_{k=1}^KV_k\times V_k$, we may have $\breve\Sigma_{O^c}\neq \tilde\Sigma_{O^c}$. For example, suppose $V_1\cap (\cup_{j>1} V_j)=\emptyset$. Then, for any orthogonal matrix $R$, the solution $(\breve\Lambda,\breve\Psi)$ given by $\breve\Lambda_{V_1} = \tilde\Lambda_{V_1} R$ and $\breve\Lambda_{V_1^c}=\tilde\Lambda_{V_1^c} $  satisfies Equation~\eqref{eq:mleequiv1} but yields $\breve\Sigma_{V_1V_1^c} = \tilde\Lambda_{V_1}R\tilde\Lambda_{V_1^c}^\T+\tilde\Psi\neq \tilde\Lambda_{V_1}\tilde\Lambda_{V_1^c}^\T+\tilde\Psi=\tilde\Sigma_{V_1V_1^c}$. Indeed, the sub-likelihood function relative to $V_1$ in Equation~\eqref{eq:linfalikelihood} is disconnected from all other sub-likelihoods and constitutes an independent factor model with a loading matrix estimate of $\Lambda_{V_1}$ affected by rotation invariance independently of the remaining rows of $\Lambda$. A similar issue arises even if all the sets $V_1,\ldots,V_K$ overlap with one another, but not sufficiently. 

In this section, we introduce the concept of \textit{linkage} between the $K$ sub-likelihood functions of the LINFA likelihood function (Equation~\eqref{eq:linfalikelihood}). This concept allows us to specify conditions for the identification of $\Sigma_{O^c}$ from $\Sigma^{(1)},\ldots,\Sigma^{(K)}$ via Equations~\eqref{eq:cov} and \eqref{eq:covk}, and thereby for 
the uniqueness of the MLE of $\Sigma$.

Let $E^{(m)}=[E^{(m)}_{ij}]$ be a $K\times K$ adjacency matrix, where \begin{equation}\label{eq:madj}
E^{(m)}_{ij}=I\left(|V_i\cap V_j|\ge m\right),
\end{equation}
and $I(\cdot)$ is the indicator function, so $E^{(m)}_{ij}=1$ if and only if $V_i$ and $V_j$ have at least $m$ nodes in common, and $E^{(m)}_{ij}=0$ otherwise. The adjacency matrix $E^{(m)}$ induces an undirected graph $G^{(m)}$ with $K$ nodes representing the variable sets $V_1,\ldots,V_K$, and with nodes $(i,j)$ connected by an edge if and only if $E^{(m)}_{ij}=1$. The graph $G^{(m)}$ can be used to characterize the \textit{$m$-linkage} between the sets $V_1,\ldots,V_K$:
\begin{definition}[\sc $m$-linkage]\label{def:linkage}
The sets $V_1,\ldots,V_K$ are $m$-linked if the $m$-linkage graph $G^{(m)}=(\{1,\ldots,K\},E^{(m)})$ is connected, i.e., every pair of distinct nodes $(i,j)$ is connected through at least one path in $E^{(m)}$. We shall say that the LINFA likelihood function in Equation~\eqref{eq:linfalikelihood} is $m$-linked if the sets $V_1,\ldots,V_K$ are $m$-linked.
\end{definition}

\begin{figure*}[t!]
    \centering
    \includegraphics[width=.8\textwidth]{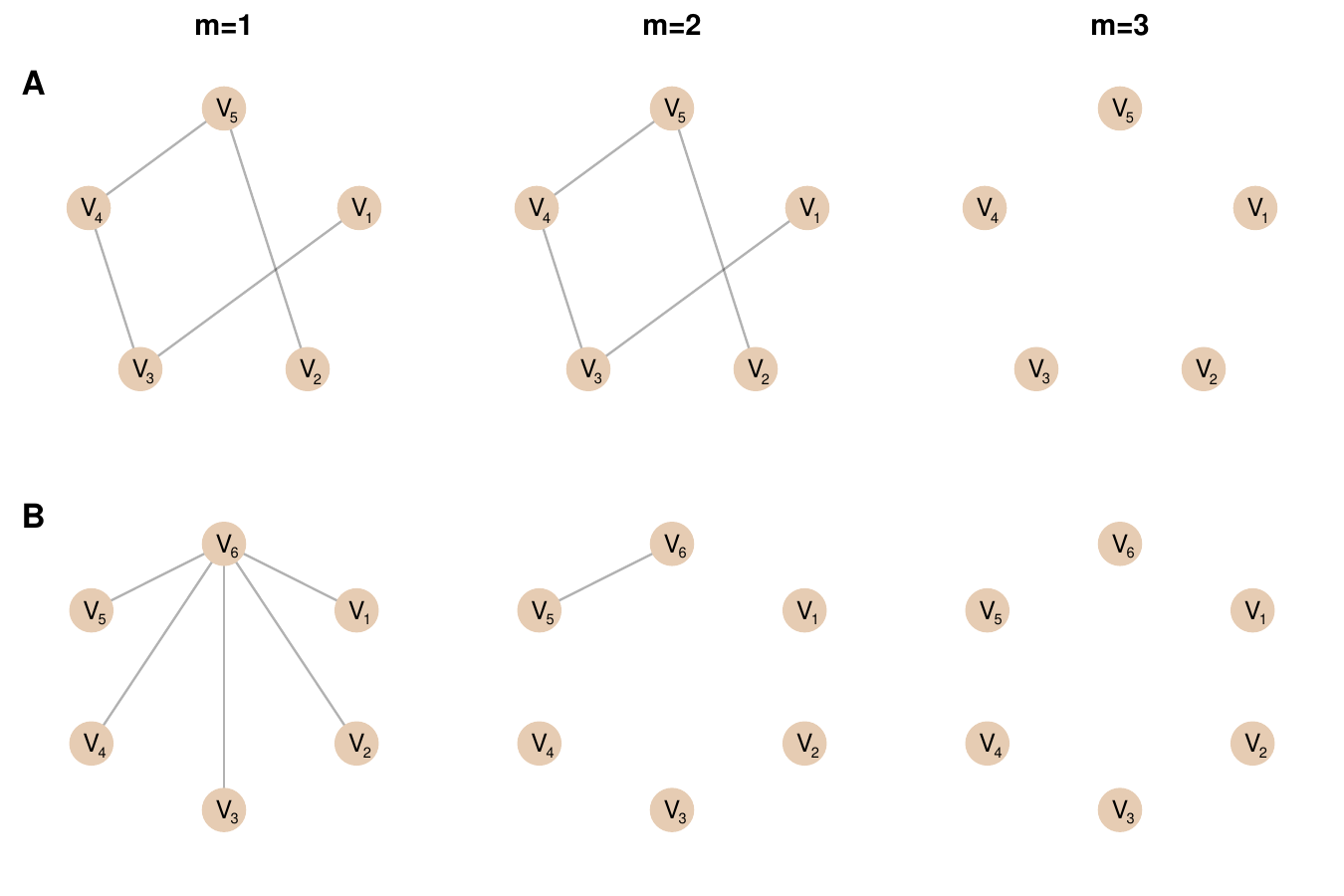}
    \caption{{\bf Linkage of vertex sets.} \textbf{(A)} Linkage graphs $G^{(m)}$ of the sets $V_1=\{1,2,3,4\}$, $V_2=\{3,4,5,6\}$, $V_3=\{5,6,7,8\}$, $V_4=\{7,8,9,10\}$, and $V_5=\{9,10,11,12\}$, for $m=1,2,3$. 
    The sets are $2$-linked but not $3$-linked. \textbf{(B)} Linkage graphs $G^{(m)}$ of the sets $V_1=\{1,2,3,4,5,6\}$, $V_2=\{1,7\}$, $V_3=\{2,8\}$, $V_4=\{3,9\}$, $V_5=\{4,5,10\}$, and $V_6=\{6,11\}$, for $m=1,2,3$. 
    The sets 
    are $1$-linked but not $2$-linked.
    }
    \label{fig:vgraph}
\end{figure*}

Thus, the sets $V_1,\ldots,V_K$ are $m$-linked if, for every pair $(V_i,V_j)$, there exists at least one sequence $V_{s_1},\ldots,V_{s_M}$ with distinct indices $1\le s_1,\ldots,s_M\le K$, $s_1=i$, $s_M=j$, and such that $V_{s_k}$ and $V_{s_{k+1}}$ have at least $m$ nodes in common for all $k=1,\ldots,M-1$. In Figure~\ref{fig:vgraph}A, we show an example with five vertex sets $V_1,\ldots,V_5$ that are 2-linked but not 3-linked; and in Figure~\ref{fig:vgraph}B, we present an example with six vertex sets $V_1,\ldots,V_6$ that are 1-linked but not 2-linked. 
We are now ready to state the condition for the identification of $\Sigma_{O^c}$ from $\Sigma_O$:

\begin{lemma}[\sc Linkage condition]\label{lemma:rotlinkage}
Let $\Sigma=\Lambda\Lambda^\T+\Psi$ and $\tilde\Sigma=\tilde\Lambda\tilde\Lambda^\T+\tilde\Psi$ be two covariance matrices, where $\Lambda,\tilde\Lambda\in\mathbb{R}^{d\times q}$, $\Psi,\tilde\Psi\in\mathcal{D}_{++}^{d\times d}$, and $1\le q<(d-1)/2$. Then, $\Sigma_{V_iV_i}=\tilde\Sigma_{V_iV_i}$, for all $i=1,\ldots,K$,  
implies $\tilde\Lambda=\Lambda R$ and $\Psi=\tilde\Psi$ (and thereby $\tilde\Sigma_{O^c}=\Sigma_{O^c}$) for some rotation matrix $R$ if and only if the sets $V_1,\ldots,V_K$ are $q$-linked.
\end{lemma}
Lemma~\ref{lemma:rotlinkage} establishes that the matrix $\Sigma$ (in particular $\Sigma_{O^c}$) can be fully identified from $\Sigma^{(1)},\ldots,\Sigma^{(K)}$ via Equations~\eqref{eq:cov} and \eqref{eq:covk} if and only if the sets $V_1,\ldots,V_K$ are $q$-linked. If the latter holds, then there exists enough overlap among the sets $V_1,\ldots,V_K$ to prevent any subset of $\tilde\Lambda$ from being arbitrarily rotated. Otherwise, arbitrary rotations may be allowed, making $\tilde\Sigma_{O^c}$ arbitrary as well. The proof of Lemma~\ref{lemma:rotlinkage} is very intensive and is presented in Appendix~\ref{app:linkage}. The importance of Lemma~\ref{lemma:rotlinkage} is reflected in the inference of $\Sigma$:
\begin{theorem}[\sc Uniqueness of the LINFA MLE]\label{theo:mlelinkage}
The LINFA MLE optimization problem in Equation~\eqref{eq:MLE} with a number of factors $1\le q<(d-1)/2$ can yield a unique solution $\hat\Sigma$ if and only if the LINFA likelihood function in Equation~\eqref{eq:linfalikelihood} is $q$-linked.
\end{theorem} 
Theorem~\ref{theo:mlelinkage} establishes that the LINFA MLE of $\Sigma$ with $q$ factors is unique if and only if the sets $V_1,\ldots,V_K$ are $q$-linked. The proof of Theorem~\ref{theo:mlelinkage} exploits Lemma~\ref{lemma:rotlinkage} and is presented in Appendix~\ref{app:linkage}.

\subsection{Expectation-Maximization Algorithm}\label{sec:EM} 
We propose an expectation-maximization (EM) algorithm for the computation of the LINFA MLE. Our algorithm can be applied to data with arbitrary missingness patterns, it involves updating equations that are all in closed form, and is accelerated by a novel \textit{Group Vertex Tessellation} (GVT) algorithm, which identifies a minimal partition of the vertex set to implement efficient maximization steps.

The EM algorithm \citep{dempster1977maximum} has also been implemented to estimate factor models in \cite{watson1983alternative}, \cite{shumway1982approach}, and \cite{banbura2014maximum}. However, \cite{watson1983alternative} assumed complete data observations, \cite{shumway1982approach} considered missing data but assumed the loading matrix $\Lambda$ to be known, and \cite{banbura2014maximum} used the Kalman filter to approximate the various conditional expectations involved in the expectation step. 

The updating equations of our LINFA MLE EM Algorithm~\ref{algo:mleEM} are derived in detail in Theorem~\ref{theo:MLEalgoUpEq}, Appendix~\ref{app:EM}, and are outlined in this section.

\subsubsection{Complete log-likelihood function}
The complete log-likelihood function for the observed data $\mathbf{X}^{(1)},\ldots,\mathbf{X}^{(K)}$ and the latent factors $\mathbf{Z}^{(1)},\ldots,\mathbf{Z}^{(K)}$ is 
\begin{eqnarray}\label{eq:complik}
&& \ell(\Lambda,\Psi; \{\mathbf{Z}^{(k)}\}_{k=1}^K, \{\mathbf{X}^{(k)}\}_{k=1}^K) = \sum_{k=1}^K \log L_k(\Lambda^{(k)},\Psi^{(k)}; \mathbf{Z}^{(k)}, \mathbf{X}^{(k)})\\
&& = -\frac{1}{2}\sum_{k=1}^K  \left\{n_k\log\det\Psi^{(k)}+{\rm tr}\left((\mathbf{X}^{(k)}-\mathbf{Z}^{(k)}\Lambda^{(k)\T})\Psi^{(k)-1}(\mathbf{X}^{(k)}-\mathbf{Z}^{(k)}\Lambda^{(k)\T})^\T\right)\right\}+C,\nonumber
\end{eqnarray}
where $C$ is a constant term that does not involve the parameters of interest $(\Lambda,\Psi)$. Note that, in our framework, the unobserved $X$'s are not treated as latent variables; thus, they are not involved in the definition of the complete likelihood function.

\subsubsection{E-step} The expectation of the complete log-likelihood function in Equation~\eqref{eq:complik} given the data $\{\mathbf{X}^{(k)}\}_{k=1}^K$ and the current estimate $(\Lambda_t,\Psi_t)$ is
\begin{eqnarray}\label{eq:Qfun}
Q_t(\Lambda,\Psi) &=& \E_t\left[\ell(\Lambda,\Psi; \{\mathbf{Z}^{(k)}\}_{k=1}^K, \{\mathbf{X}^{(k)}\}_{k=1}^K)\mid \{\mathbf{X}^{(k)}\}_{k=1}^K\right]\\
& = & -\frac{1}{2}\sum_{k=1}^K   \Big\{n_k\log\det \Psi^{(k)}+{\rm tr}\left(\mathbf{X}^{(k)\T}\mathbf{X}^{(k)}\Psi^{(k)-1} \right) \nonumber \\ 
& &  +{\rm tr}\left(S_t^{(k)}\Lambda^{(k)\T}\Psi^{(k)-1}\Lambda^{(k)}\right) -2{\rm tr}\left(M_t^{(k)}\Lambda^{(k)\T}\Psi^{(k)-1}\mathbf{X}^{(k)\T} \right)\Big\}+C,\nonumber
\end{eqnarray}
where $\E_t[*]$ denotes expectation, assuming the distribution of $\mathbf{X}^{(1)},\ldots,\mathbf{X}^{(K)}$ has parameters equal to $(\Lambda_t,\Psi_t)$, and
\begin{eqnarray}
    M_t^{(k)} &=& \mathbf{X}^{(k)}\Gamma^{(k)}_t \label{eq:Mt}\\
    S_t^{(k)} &=& n_k(I_q-\Gamma^{(k)\T}_t\Lambda^{(k)}_t)+M^{(k)\T}_t M_t^{(k)}\label{eq:St}\\
    \Gamma^{(k)}_t &=& A^{(k)}_t(I_q-(I_q+B^{(k)}_t)^{-1}B^{(k)}_t)\label{eq:Gt}\\
    A^{(k)}_t &=& \Psi^{(k)-1}_t\Lambda^{(k)}_t\label{eq:At}\\
    B^{(k)}_t &=& A_t^\T\Lambda^{(k)}_t\label{eq:Bt}
\end{eqnarray}

\subsubsection{M-step accelerated by Group Vertex Tessellation}\label{sec:Mstep} 
The expected complete loglikelihood function $Q_t(\Lambda,\Psi)$ in Equation~\eqref{eq:Qfun} could be maximized with respect to all parameters relative to node $i$, i.e., $\Lambda_{i1},\ldots,\Lambda_{iq},\Psi_{ii}$, through repeated cycles of iterations for $i=1,\ldots,d$. In each iteration, all samples relative to node $i$ would be involved. However, this procedure would require $d$ computations per cycle. This computational burden can be substantially reduced with the Group Vertex Tessellation (GVT) Algorithm~\ref{algo:gvt}. This algorithm allows us to identify a minimal partition $W_1,\ldots,W_J$ of the vertex set $V=\{1,\ldots,d\}$, where $J\le d$ and all nodes in $W_k$ are observed simultaneously on exactly the same samples across the $K$ data sets $\mathbf{X}^{(1)},\ldots,\mathbf{X}^{(K)}$, enabling us to implement the M-step with closed form expressions and the smallest possible number of operations. 

\begin{figure*}[t!]
\centering
\includegraphics[width=1\textwidth]{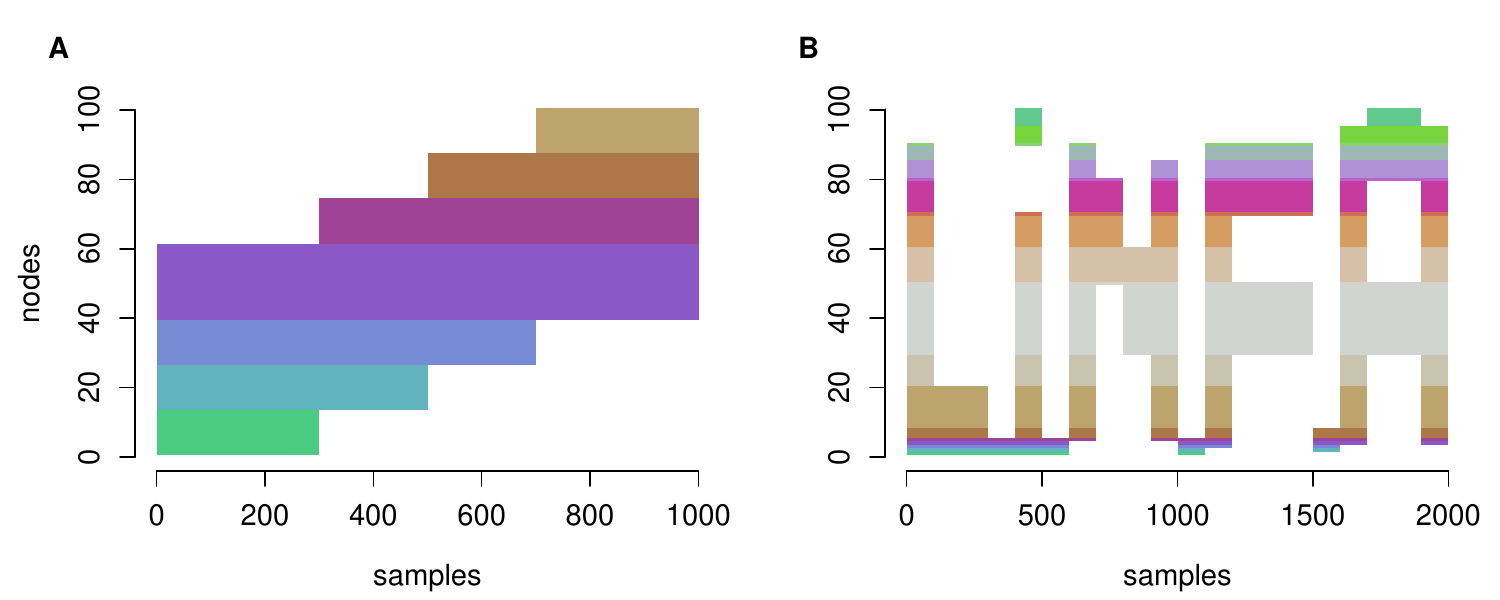}
\caption{\textbf{Group Vertex Tessellation}. \textbf{(A)} Vertex tessellation in a  sequential observation pattern. The node subsets $W_1,\ldots,W_J$ produced by Algorithm~\ref{algo:gvt} are denoted by different colors.  \textbf{(B)} Vertex tessellation in a nonsequential observation pattern.}\label{fig:gvt}
\end{figure*}

\begin{algorithm}[ht!]\small
{\bf Input:}  $V_1,\ldots,V_K$, where $\cup_{k=1}^K V_k=V=\{1,\ldots,d\}$.
\begin{enumerate}
\item Create $I\in\{0,1\}^{d\times K}$, where $I_{ik}=1$ if $i\in V_k$, and $I_{ik}=0$ otherwise. 
\item Compute the distance matrix $\Delta\in\mathbb{R}_+^{d\times d}$, where $\Delta_{ij}=\Vert I_{i.}-I_{j.} \Vert_1 $.
\item Obtain the set 
$\mathcal{W}=\big\{\{j\in V:\Delta_{ij}=0\}: i\in V\big\}$.
\end{enumerate}
{\bf Output:} $\mathcal{W}=\{W_1,\ldots,W_J\}$.
\caption{{\sc Group Vertex Tessellation (GVT)}}\label{algo:gvt}
\end{algorithm}
Figure~\ref{fig:gvt}A shows a case where $V_1=\{1,\ldots,61\}$, $V_2=\{14,\ldots,74\}$, $V_3=\{27,\ldots,87\}$, and $V_4=\{40,\ldots,100\}$, and the GVT algorithm yields the partition $W_1=\{1,\ldots,13\}$, $W_2=\{14,\ldots,26\}$, $W_3=\{27,\ldots,39\}$, $W_4=\{40,\ldots,61\}$, $W_5=\{62,\ldots,74\}$, $W_6=\{75,\ldots,87\}$, and $W_7=\{88,\ldots,100\}$.  Figure~\ref{fig:gvt}B shows a case of non-serial overlap. 

The optimality of the GVT Algorithm~\ref{algo:gvt} is established in the following theorem:
\begin{theorem}[\sc Optimality of the GVT Algorithm~\ref{algo:gvt}]\label{theo:gvt}
Let $V_1,\ldots,V_K \subseteq V=\{1,\ldots, d\}$ be such that $\cup_{k=1}^K V_k=V$. Define the sets $\mathcal{A}_O=\{(i,k):~i\in V_k,~i=1,\ldots,d,~k=1,\ldots,K\}$ and $\mathcal{K}_U:=\left\{k:~U\subseteq V_k\right\}$. 
Then, the collection $W_1,\ldots,W_J$ produced by the GVT Algorithm~\ref{algo:gvt}  is (i) a partition of $V$ and (ii) the solution of the optimization problem
\begin{equation}\label{eq:gvteq}
\begin{array}{l}
{\rm minimize} ~~J\\
{\rm s.t.}~~~ \mathcal{A}_O = \bigcup\limits_{j=1}^J \left(W_j\times \mathcal{K}_{W_j}\right)
\end{array}
\end{equation}
\end{theorem}
Theorem~\ref{theo:gvt} establishes that the GVT Algorithm~\ref{algo:gvt} allows us to partition the observed data portion ($\mathcal{A}_O$) into the smallest number of Cartesian sets $W_1\times \mathcal{K}_{W_1}$, $\ldots$, $W_J\times \mathcal{K}_{W_J}$. Hence, maximizing the function $Q(\Lambda,\Psi)$ (Equation~\eqref{eq:Qfun}) with respect to parameter portions relative to the variable sets $W_1,\ldots W_J$ is optimal. The proof of Theorem~\ref{theo:gvt} is in Appendix~\ref{app:EM}. 

We now show that the expected complete log-likelihood function $Q_t(\Lambda,\Psi)$ (Equation~\eqref{eq:Qfun}) can be maximized sequentially with respect to $(\Lambda_{W_j},\Psi_{W_j})$, for $j=1,\ldots,J$. 
Let $\mathcal{K}_{W_j}=\{k:W_j\subseteq V_k\}$ be the set of indices of the data sets where all nodes in $W_j$ were fully observed. The gradient of $Q_t$ with respect to the portion $\Lambda_{W_j}=[\Lambda_{il}]_{i\in W_j,l\in\{1,\ldots,q\}}$ is
\begin{eqnarray*}
\frac{\partial Q_t}{\partial\Lambda_{W_j}} &=& \Psi^{-1}_{{W_j}{W_j}}\left(-\Lambda_{W_j}\sum_{k\in\mathcal{K}_{W_j}}S^{(k)}_t 
+\sum_{k\in\mathcal{K}_{W_j}}\mathbf{X}^{(k)\T}_{W_j}M_t^{(k)}\right)
\end{eqnarray*} 
and, setting it equal to zero, we obtain the updating equation
\begin{equation}\label{eq:upLamb}
\Lambda_{{W_j},t+1} = \left(\sum_{k\in\mathcal{K}_{W_j}}\mathbf{X}^{(k)\T}_{W_j} M_t^{(k)}\right)\left(\sum_{k\in\mathcal{K}_{W_j}}S^{(k)}_t \right)^{-1}
\end{equation}
where $\sum_{k\in\mathcal{K}_{W_j}}S^{(k)}_t$ is guaranteed to be positive definite, hence invertible, by  Lemma~\ref{lemma:posdefSt} in Appendix~\ref{app:lemmas}. 
Applying Equation~\eqref{eq:upLamb} for all $j=1,\ldots,J$ yields the updated $\Lambda_{t+1}$. 

Next, let $\Omega=\Psi^{-1}$. The gradient of $Q_t$ with respect to the portion $\Omega_{W_jW_j}$ is
\begin{eqnarray*}
\frac{\partial Q_t}{\partial \Omega_{{W_j}{W_j}}} &=& \frac{1}{2}\sum_{k\in\mathcal{K}_{W_j}}\left(n_k\Omega_{{W_j}{W_j}}^{-1} - \mathbf{X}^{(k)\T}_{W_j}\mathbf{X}^{(k)}_{W_j} -\Lambda_{{W_j}}S_t^{(k)}\Lambda^\T_{{W_j}} +2\mathbf{X}^{(k)\T}_{W_j}M_t^{(k)}\Lambda^{T}_{{W_j}}\right)
\end{eqnarray*}
By setting this gradient to zero and then plugging $\Lambda_{W_j,t+1}$ in place of $\Lambda_{W_j}$ (Equation~\eqref{eq:upLamb}), we obtain the updating equation
\begin{equation}\label{eq:upPsi}
\Psi_{{W_j}{W_j},t+1} = {\rm Diag}\left\{\sum\limits_{k\in\mathcal{K}_{W_j}}\left(\mathbf{X}_{W_j}^{(k)\T}\mathbf{X}^{(k)}_{W_j} -\Lambda_{{W_j},t+1}S^{(k)}_t\Lambda_{{W_j},t+1}^\T \right)\right\}\left(\sum\limits_{k\in\mathcal{K}_{W_j}}n_k\right)^{-1},
\end{equation}
where ${\rm Diag}(A)$ denotes the diagonal matrix with the same diagonal entries as the square matrix $A$ (i.e., ${\rm Diag}(A) = A\odot I$, where $I$ is the identity matrix and $\odot$ denotes the Hadamard product). Applying Equation~\eqref{eq:upPsi} for all $j=1,\ldots,J$ yields the updated $\Psi_{t+1}$. 
The solution $(\Lambda_{t+1},\Psi_{t+1})$ is a maximum point since the Hessian matrix of $Q_t$ evaluated at $(\Lambda_{t +1}, \Psi_{t +1})$ is negative definite, as shown in the proof of Theorem~\ref{theo:MLEalgoUpEq} in Appendix~\ref{app:EM}. 

\begin{algorithm}[ht!]
{\bf Input}: Data $\mathbf{X}^{(1)},\ldots,\mathbf{X}^{(K)}$; $V_1,\ldots,V_K$; start values $\Lambda\in\mathbb{R}^{d\times q}$ and $\Psi\in\mathcal{D}_{++}^{d\times d}$\;
\begin{enumerate}
\item Obtain $W_1,\ldots,W_J$ from $V_1,\ldots,V_K$ via Algorithm~\ref{algo:gvt}, and $D_{W_1},\ldots,D_{W_J}$, where 
\[
D_W = \frac{1}{n_W}{\rm Diag}\left(\sum_{k\in\mathcal{K}_W}\mathbf{X}_W^{(k)\T}\mathbf{X}_W^{(k)}\right),
\]
$\mathcal{K}_W = \{k:W\subseteq V_k \} $, and $n_W = \sum_{k\in\mathcal{K}_W}n_k$.
\item Iterate until convergence:
\begin{enumerate}
\item[] {\sc E-step.}~ For $k=1,\ldots,K$, update
\begin{eqnarray*}
A^{(k)} &=& \Psi^{(k)-1}\Lambda^{(k)}\\
B^{(k)} &=& A^{(k)\T}\Lambda^{(k)}\\
\Gamma^{(k)} & = & A^{(k)}(I_q-(I_q+B^{(k)})^{-1}B^{(k)})\\
M^{(k)} &=& \mathbf{X}^{(k)}\Gamma^{(k)}\\
S^{(k)} &=&  n_k(I_q-\Gamma^{(k)\T}\Lambda^{(k)})+M^{(k)\T} M^{(k)}
\end{eqnarray*}
\item[] {\sc M-step.}~ For $j=1,\ldots,J$, compute $S_{W_j} = \sum_{k\in\mathcal{K}_{W_j}}S^{(k)}$ and update
\begin{eqnarray*}
\Lambda_{W_j} &=&  \left(\sum_{k\in\mathcal{K}_{W_j}}\mathbf{X}^{(k)\T}_{W_j} M^{(k)}\right)S_{W_j}^{-1}\\
\Psi_{{W_j}} &=& 
D_{W_j}
-\frac{1}{n_{W_j}}{\rm Diag}\left(\Lambda_{{W_j}}S_{W_j}\Lambda_{{W_j}}^\T\right),
\end{eqnarray*}
\end{enumerate}
\item Rotation: obtain $\hat\Lambda = \Lambda R$, where $R$ is the matrix of eigenvectors of $\Lambda^\T\Psi^{-1}\Lambda$.
\item Sign flip: update $\hat\Lambda_{.j}\equiv \hat\Lambda_{.j}\cdot {\rm sign}(\hat\Lambda_{jj}), ~j=1,\ldots,q$.
\end{enumerate}
{\bf Output}: Maximum likelihood estimate $(\hat\Lambda, \hat\Psi)$.
\caption{LINFA MLE EM Algorithm}\label{algo:mleEM}
\end{algorithm}

\subsubsection{Full algorithm}
Iterating the E and M steps produces a sequence of solutions $(\Lambda_0,\Psi_0)$, $(\Lambda_1,\Psi_1)$, \ldots that will converge to one of infinitely many possible solutions of the MLE problem in Equation~\eqref{eq:MLE}. Applying the rotation 
\begin{equation}\label{eq:rotMLE}
\hat\Lambda \equiv \hat\Lambda R,
\end{equation}
where $R$ is the matrix of eigenvectors of $\hat\Lambda^\T\hat\Psi^{-1}\hat\Lambda$ \citep{rao1955estimation,jennrich1974simplified}, ensures the uniqueness of $\hat\Lambda$ up to a column sign flip as per Equation~\eqref{eq:canon}, provided that the linkage condition established in Theorem~\ref{lemma:rotlinkage} holds. Note that this rotation has no effect on the estimate $\hat\Sigma$ (Lemma~\ref{lemma:canonrot} in Appendix~\ref{app:lemmas}). Moreover, we apply a column sign correction
\begin{equation}\label{eq:signflip}
\hat\Lambda_{.j}\equiv \hat\Lambda_{.j}\cdot {\rm sign}(\hat\Lambda_{jj}), ~j=1,\ldots,q, 
\end{equation}
which yields a unique MLE $(\hat\Lambda,\hat\Psi)$, where the entries in the first row of $\hat\Lambda$ are all positive. Note that $\hat\Lambda$ is dense almost surely, and so ${\rm sign}(\hat\Lambda_{jj})$ is either $-1$ or $1$ almost surely.

The full EM optimization procedure is summarized in Algorithm~\ref{algo:mleEM}, where the steps are arranged to minimize redundant computations, thereby reducing computer memory pressure and processor load. Algorithm~\ref{algo:mleEM} takes as input the observed data $\mathbf{X}^{(1)},\ldots,\mathbf{X}^{(K)}$, the corresponding observed node subsets $V_1,\ldots,V_K$, and the starting values for $\Lambda\in\mathbb{R}^{d\times q}$ and $\Psi\in\mathcal{D}_{++}^{d\times d}$ (see Appendix~\ref{app:startvalues}). 
In step 1, the GVT Algorithm~\ref{algo:gvt} yields the vertex partition $W_1,\ldots,W_J$, and we compute the diagonal matrices $D_{W_1},\ldots,D_{W_J}$, which are constants to be used throughout all EM iterations. In step 2, the EM steps are iterated until convergence (e.g., relative difference between log-likelihoods at two consecutive parameter estimate updates is smaller than a given threshold). In step 3, the factor loadings are rotated as per Equation~\eqref{eq:rotMLE}, and finally, in step 4, the signs of the columns of the loading matrix are flipped as per Equation~\eqref{eq:signflip}. 
The algorithmic convergence of the EM Algorithm~\ref{algo:mleEM} is established in the following theorem:
\begin{theorem}[\sc Convergence of the LINFA MLE EM Algorithm~\ref{algo:mleEM}]\label{theo:MLEalgoconv}
Any converging solution of the LINFA MLE EM Algorithm~\ref{algo:mleEM} satisfies the first order condition of the MLE optimization problem in Equation~\eqref{eq:MLE}.
\end{theorem}
The proof of Theorem~\ref{theo:MLEalgoconv} is very intensive and is presented in Appendix~\ref{app:EM}.

\subsection{Statistical properties of the LINFA MLE}\label{sec:theory} In this section, we establish the consistency and asymptotic normality of the LINFA MLE. In Section~\ref{sec:ass}, we state our assumptions. In Section~\ref{sec:mainres}, we present our two main theorems on the consistency and asymptotic normality of the LINFA MLE. Finally, in Section~\ref{sec:uncertainty}, we propose confidence regions, hypothesis tests, and bootstrap methods.

\subsubsection{Assumptions}\label{sec:ass}
Consider the following assumptions:
\begin{enumerate}
    \item[] \textbf{A1}: The sets $V_1,\ldots,V_K$ are $q$-linked, where $1\le q<(d-1)/2$.
    
    \item[] \textbf{A2}: There exist $p_1,\ldots,p_k\in(0,1)$ such that $\sum_{k=1}^Kp_k=1$ and $n_k=\lfloor n\cdot p_k\rceil$, for all $k=1,\ldots,K$.
    
    \item[] \textbf{A3}: $(\Lambda,\Psi)\in\mathcal{P}_b$, where
    \begin{eqnarray}\nonumber
       \mathcal{P}_b&:=& \big\{(\Lambda,\Psi)\in\mathbb{R}^{d\times q}\times \mathcal{D}^{d\times d}_{++}: \Lambda^\T\Psi^{-1}\Lambda=D\in \mathcal{D}^{d\times d}_{++}, 0<D_{qq}<\ldots<D_{11}<\infty,\\
        &&
        ~\min_{1\le i\le d}\Psi_{ii}\ge b, ~\Vert\Lambda\Vert_F^2+\Vert\Psi\Vert_\infty\le b^{-1}, \Lambda_{jj}>0,\forall j=1,\ldots,q\big\}\label{eq:theoset}
    \end{eqnarray}
    for some $b\in(0,1)$.
    
    \item[] \textbf{A4}: The MLE $(\hat\Lambda,\hat\Psi)$  of $(\Lambda,\Psi)$ is constrained within the set $\mathcal{P}_b$.
\end{enumerate}

Assumption A1 requires the observed variable sets to be $q$-linked to allow for the identification of $\Sigma_{O^c}$ from $\Sigma_O$ (Lemma~\ref{lemma:rotlinkage}). Assumption A2 requires that the sample sizes $n_1,\ldots,n_K$ of the $K$ data sets increase with $n$ according to the proportions $p_1,\ldots,p_K$. Thus, $n\to\infty$ implies $n_j\to\infty$ and $\frac{n_j}{n}\to p_j$, for all $j=1,\ldots, K$. Assumption A3 requires that the true parameter values $(\Lambda,\Psi)\in\mathcal{P}_b$ are bounded and satisfy the canonical rotation constraint (Equation~\eqref{eq:canon}), whereby the effective number of free parameters in $(\Lambda,\Psi)$ is 
\begin{equation}\label{eq:kappa}
    \kappa~=~ d(q+1)-q(q-1)/2,
\end{equation}
where $d(q+1)$ is the number of entries in $(\Lambda,\Psi)$ and $q(q-1)/2$ is the number of constraints. 
Moreover, the positivity constraint on the diagonal entries of $\Lambda$ allows for the identification of the column signs of $\Lambda$. Note that the positive constant $b$ can be taken as small as necessary to make $\mathcal{P}_b$ large enough to include specific situations of interest. For example, if we rather assumed $0<c_1\le \Psi_{ii}\le c_2 $ and $\Vert\Lambda\Vert_\infty\le c_3<\infty$, then we would just need $0<b\le\min\{c_1,(c_2+dqc_3^2)^{-1}\}$. Finally, Assumption A4 simply requires the MLE to be constrained within the same set as the true value of the parameter.

\subsubsection{Consistency and asymptotic normality}\label{sec:mainres}
The following two theorems establish the consistency and asymptotic normality of the LINFA MLE, respectively, in terms of $\theta = ({\rm vec}(\Lambda)^\T,{\rm diag}(\Psi)^\T)^\T$ and $\hat\theta = ({\rm vec}(\hat\Lambda)^\T,{\rm diag}(\hat\Psi)^\T)^\T$:
\begin{theorem}[\sc Consistency]\label{theo:consistency}
Under assumptions A1--A4, as $n\to\infty$,
\begin{eqnarray}
\Vert \hat\theta-\theta\Vert &\pto& 0
\end{eqnarray}
where $\Vert\cdot\Vert$ is any vector norm.
\end{theorem}

In the proof of Theorem~\ref{theo:consistency} in Appendix~\ref{app:proofstheory}, we first show that $\hat\Lambda$ is a consistent estimator of $\Lambda$ up to column sign flips. Then, due to A3, we apply the column sign corrections as per Equation~\eqref{eq:signflip}, ensuring $\hat\Lambda_{jj}>0$ for all $j=1,..,q$, and thereby allowing for the exact consistency of $\hat\theta$, which we need to prove the asymptotic normality of the LINFA MLE:

\begin{theorem}[\sc Asymptotic Normality]\label{theo:asynorm}
Under assumptions A1--A4, as $n\to\infty$, 
\begin{eqnarray}\label{eq:asnorm}
    \sqrt{n}(\hat\theta-\theta) &\dto& N(0,\Omega),
\end{eqnarray}
where:
\begin{eqnarray}
\Omega &=& A\tilde{\mathcal{I}}(\theta)A^\T\label{eq:acov}
\end{eqnarray}
is the asymptotic covariance matrix;
\begin{eqnarray}\label{eq:fishinfo}
    \tilde{\mathcal{I}}(\theta) &=&\sum\limits_{t=1}^K p_t \tilde{\mathcal{I}}_t(\theta)
\end{eqnarray}
is the unconstrained Fisher Information Matrix; 
\begin{eqnarray}\label{eq:fishinfok}
\tilde{\mathcal{I}}_t(\theta) &=& -\E\left[\tfrac{\partial^2}{\partial\theta\partial\theta^\T}\log f_{X_{V_t}}(X_{V_t};\theta^{(t)})  \right]
\end{eqnarray}
has entries equal to
\begin{eqnarray}
\tilde{\mathcal{I}}_{t,(\Lambda_{ij},\Lambda_{kl})}(\theta) &=& (\Sigma^{-1}_{V_tV_t}\Lambda_{V_t})_{ik}\cdot I(k=l)+(\Sigma^{-1}_{V_tV_t}\Lambda_{V_t})_{jk}\cdot I(i=l)\\
\tilde{\mathcal{I}}_{t,(\Psi_{ii},\Psi_{kk})}(\theta)&=&\frac{1}{2}([\Sigma^{-1}_{V_tV_t}]_{ik})^2\\
\tilde{\mathcal{I}}_{t,(\Lambda_{ij},\Psi_{kk})}(\theta)&=&(\Sigma^{-1}_{V_tV_t}\Lambda_{V_t})_{ik}(\Sigma_{V_tV_t}^{-1})_{jk},
\end{eqnarray}
for $ i,k\in V_t$, and equal to zero otherwise; 
\begin{eqnarray}
    A &=& [\mathcal{I}(\theta)^{-1}]_{1:d(q+1),1:d(q+1)},\label{eq:Amatrix}\end{eqnarray}
where
\begin{eqnarray}
    \mathcal{I}(\theta) &=& \left[ \begin{array}{cc}
\tilde{\mathcal{I}}(\theta), & g'(\theta)^\T\\
g'(\theta), & 0
    \end{array}
\right]  \label{eq:augfishinfo}
\end{eqnarray}
is the augmented Fisher Information Matrix; and  $g(\theta)=\left(g_{ij}(\theta)\right)_{1\le i<j\le q}$ has entry
\begin{eqnarray}
g_{ij}(\theta) &=& [\Lambda^\T\Psi^{-1}\Lambda]_{ij} =  \sum_{h=1}^d\Lambda_{hi}\Lambda_{hj}/\Psi_{hh}
\end{eqnarray}
\end{theorem}
The \textit{unconstrained Fisher Information Matrix} $\tilde{\mathcal{I}}(\theta)$ in Equation~\eqref{eq:fishinfo} is the weighted average of the Fisher information matrices $\tilde{\mathcal{I}}_1(\theta),\ldots,\tilde{\mathcal{I}}_K(\theta) $ (Equation~\eqref{eq:fishinfok}) about $\theta$ contained in the $K$ data blocks, without assuming the canonical rotation constraint in Equation~\eqref{eq:canon}.  
Thus, the matrix $\tilde{\mathcal{I}}(\theta)$ is not invertible and cannot be used directly to establish the asymptotic normality of the LINFA MLE, which is rotated via Equation~\eqref{eq:rotMLE}. The \textit{augmented Fisher information matrix} $\mathcal{I}(\theta)$ in Equation~\eqref{eq:augfishinfo} contains the gradient component $g'(\theta)$ that allows us to impose the canonical rotation constraint in Equation~\eqref{eq:canon} implied by $g(\theta)=0$. This augmentation makes $\mathcal{I}(\theta)$ invertible, and the upper-left $d(q+1)\times d(q+1)$ portion of its inverse, the matrix $A$ in Equation~\eqref{eq:Amatrix}, is then used to find the asymptotic covariance matrix $\Omega$ in Equation~\eqref{eq:acov}. The sign flip correction in Equation~\eqref{eq:signflip} allows for $\hat\theta-\theta\pto 0$ in Equation~\eqref{eq:asnorm} as per Theorem~\ref{theo:consistency}, which is necessary for asymptotic normality. The complete proof of Theorem~\ref{theo:asynorm} is in Appendix~\ref{app:proofstheory}.

The following corollary is a consequence of Theorems~\ref{theo:consistency} and \ref{theo:asynorm}, the Continuous Mapping Theorem, and the Multivariate Delta Method:
\begin{corollary}\label{coro:sigma} Under assumptions A1-A4, as $n\to\infty$,
    \begin{eqnarray}
        \Vert{\rm vec}(\hat\Sigma-\Sigma)\Vert&\pto& 0\label{eq:consistSigma}\\\label{eq:acovsigma}
        \sqrt{n}~ {\rm vec}(\hat\Sigma-\Sigma)&\dto& N(0,J\Omega J^\T),
    \end{eqnarray}
    where $\Vert*\Vert$ is any vector norm, 
  $\Omega$ is given in Equation~\eqref{eq:acov}, $J = [\Lambda\otimes I_d+I_d\otimes\Lambda\times C^{(d,q)},S]$, 
$C^{(d,q)}$ is a commutation matrix, $S$ is a $d^2\times d^2$ matrix with the $j$-th column equal to ${\rm vec}(E^{(j)})$,  $E^{(j)}$ is a matrix with entry $ E^{(j)}_{lk}=I(l=k=j)$, and $I()$ is the indicator function.
\end{corollary}

\subsubsection{Confidence regions and hypothesis testing}\label{sec:uncertainty}
The asymptotic results established in Theorem~\ref{theo:asynorm} can be used to approximate estimates of the covariance matrix of the LINFA MLE $\hat\theta$ as $\hat\Omega/n$, where $\hat\Omega$ is the MLE of the covariance matrix $\Omega$ in Equation~\eqref{eq:acov}. Moreover, confidence regions and hypothesis tests about the parameter vector $\theta$ can be constructed via the likelihood-ratio approach. In Theorem~\ref{theo:confreg}, we propose a confidence region with approximate coverage $1-\alpha$: 
\begin{theorem}[\sc Confidence Region]\label{theo:confreg} Let $\ell_n(\Lambda,\Psi)$ be the logarithm of the LINFA likelihood function in Equation~\eqref{eq:linfalikelihood}, and let $(\hat\Lambda,\hat\Psi)$ be the LINFA MLE of $(\Lambda,\Psi)$. Under assumptions A1--A4, for large $n$, an approximate $1-\alpha$ confidence region for $(\Lambda,\Psi)$ is 
    \begin{equation}\label{eq:confreg}
        \mathcal{C}_{1-\alpha}~=~\left\{(\Lambda,\Psi)\in\mathcal{P}_b: ~
        \lambda_n\le \chi^2_{\kappa,\alpha}\right\},
    \end{equation}
    where $\mathcal{P}_b$ is the set in Equation~\eqref{eq:theoset}, $\lambda_n:=2(\ell_n(\hat\Lambda,\hat\Psi)-\ell_n(\Lambda,\Psi))$, and $\chi^2_{\kappa,\alpha}$ is the $1-\alpha$ quantile of the $\chi^2_\kappa$ distribution with degrees of freedom $\kappa$ given in  Equation~\eqref{eq:kappa}. 
\end{theorem} Theorem~\ref{theo:confreg} is proved in Appendix~\ref{app:proofstheory}. Based on the confidence region in Equation~\eqref{eq:confreg}, a simple hypothesis test can also be established: 
\begin{corollary}[\sc Hypothesis Test]\label{coro:ht}
 Let $(\Lambda_0,\Psi_0)\in\mathcal{P}_b$. Consider testing the hypotheses $H_0:(\Lambda,\Psi)=(\Lambda_0,\Psi_0)$ versus $H_1:(\Lambda,\Psi)\in\mathcal{P}_b\setminus \{(\Lambda_0,\Psi_0)\}$.  
Under assumptions of Theorem~\ref{theo:confreg}, rejecting $H_0$ if $(\Lambda_0,\Psi_0)\notin\mathcal{C}_{1-\alpha}$ controls the type I error at level $\alpha$ asymptotically.
\end{corollary}
The standard errors of factor model estimators can also be approximated via various resampling techniques, as described in \cite{zhang2014estimating}. In this paper, we consider two bootstrap approaches \citep{efron1994introduction}, adapting them to our framework of structural missingness: the \textit{parametric bootstrap} and the \textit{nonparametric bootstrap}. 

\begin{algorithm}[ht!]
{\bf Input}: MLE $(\hat\Lambda,\hat \Psi)$; $V_1,\ldots,V_K$; sample sizes $n_1,\ldots,n_K$ of the $K$ observed data $\mathbf{X}^{(1)},\ldots,\mathbf{X}^{(K)}$; real valued function  $\tau=g(\Lambda,\Psi)$; number of draws $B$\;

For $b=1,\ldots,B$:
\begin{enumerate}
    \item Generate $K$ independent datasets $\tilde{\mathbf{X}}^{(1)},\ldots,\tilde{\mathbf{X}}^{(K)}$, where $\tilde{\mathbf{X}}_k$ consists of $n_k$ i.i.d. samples drawn from $N(0,\hat\Lambda^{(k)}\hat\Lambda^{(k)\T}+\hat\Psi^{(k)})$.
    \item Compute the MLE $(\tilde\Lambda(b),\tilde\Psi(b))$ via Algorithm~\ref{algo:mleEM} based on $\tilde{\mathbf{X}}^{(1)},\ldots,\tilde{\mathbf{X}}^{(K)}$.
    \item Compute $\tilde\tau_b=g(\tilde\Lambda(b),\tilde\Psi(b))$.
\end{enumerate}
{\bf Output}: Estimate of the standard error of the MLE $\hat\tau$:
\[
\widehat{SE}(\hat\tau) = \sqrt{\tfrac{1}{B-1}\sum_{b=1}^B \left(\tilde\tau_b-\tfrac{1}{B}\sum_{s=1}^B\tilde\tau_s\right)^2}
\]
\caption{LINFA Parametric bootstrap}\label{algo:parbootstrap}
\end{algorithm}

\begin{algorithm}[ht!]
{\bf Input}: Data sets $\mathbf{X}^{(1)},\ldots,\mathbf{X}^{(K)}$ with sample sizes $n_1,\ldots,n_K$, respectively; $V_1,\ldots,V_K$; real valued function $\tau=g(\Lambda,\Psi)$; number of draws $B$\;

For $b=1,\ldots,B$:
\begin{enumerate}
    \item Generate $K$ independent datasets $\tilde{\mathbf{X}}_1,\ldots,\tilde{\mathbf{X}}_K$, where $\tilde{\mathbf{X}}_k$ consists of $n_k$ samples drawn with replacement from $\mathbf{X}^{(k)}$.
    \item Compute the MLE $(\tilde\Lambda(b),\tilde\Psi(b))$ via Algorithm~\ref{algo:mleEM} based on $\tilde{\mathbf{X}}_1,\ldots,\tilde{\mathbf{X}}_K$.
    \item Compute $\tilde\tau_b=g(\tilde\Lambda(b),\tilde\Psi(b))$.
\end{enumerate}
{\bf Output}: Estimate of the standard error of the MLE $\hat\tau$:
\[
\widehat{SE}(\hat\tau) = \sqrt{\tfrac{1}{B-1}\sum_{b=1}^B \left(\tilde\tau_b-\tfrac{1}{B}\sum_{s=1}^B\tilde\tau_s\right)^2}
\]
\caption{LINFA Nonparametric bootstrap}\label{algo:nonparbootstrap}
\end{algorithm}

Suppose we want to approximate the standard error of the MLE $\hat\tau=g(\hat\Lambda,\hat\Psi)$ of a function $\tau=g(\Lambda,\Psi)$, for example, $\tau:=\Sigma_{ij}=\sum_{l=1}^q\Lambda_{il}\Lambda_{jl}+\Psi_{ij}$.  The parametric bootstrap (Algorithm~\ref{algo:parbootstrap}) approximates the standard error of $\hat\tau$ with the empirical standard deviation of multiple MLEs $\tilde\tau_1,\ldots,\tilde\tau_B$ obtained from $B$ artificial datasets drawn from the same parametric distribution (Normal distribution) with parameters set equal to the MLE $(\hat\Lambda,\hat\Psi)$. 
The nonparametric bootstrap (Algorithm~\ref{algo:nonparbootstrap}) approximates the standard error of $\hat\tau$ in a similar way, but with the $B$ artificial datasets drawn with replacement from the original data rather than from the estimated parametric distribution. In Section~\ref{sec:simuncertainty}, we show that both parametric and nonparametric bootstrap methods allow us to adequately approximate standard errors and that they are comparable with the asymptotic approximations based on Theorem~\ref{theo:asynorm} and Corollary~\ref{coro:sigma}.

\subsection{Model selection criteria}\label{sec:modelselection}
There are several criteria to select the number of factors $q$ in factor analysis \citep{preacher2013choosing,haslbeck2022estimating}. We consider three approaches, adapting them to our framework of structural missingness: \textit{$N$-fold cross-validation} (N-CV), \textit{Akaike Information Criterion} (AIC) \citep{akaike1987factor}, and \textit{Bayesian Information Criterion} (BIC) \citep{schwarz1978estimating}. 

We implement N-CV as follows. In our observation setting, we observe $K$ independent data sets $\mathbf{X}^{(1)},\ldots,\mathbf{X}^{(K)}$. Thus, we randomly split each observed data set $\mathbf{X}^{(i)}$ into $N$ disjoint subsets $\mathbf{X}^{(i)}_1,\ldots,\mathbf{X}^{(i)}_N$ with approximately equal sample sizes. Then, we let $\mathcal{X}_j=\{ \mathbf{X}^{(1)}_j, \ldots, \mathbf{X}^{(K)}_j\}$ be the $j$-th fold, and we define the LINFA $N$-fold cross-validation risk
\begin{equation}\label{eq:cv}
\textsc{Risk}_{\rm NCV}(q)~=~-\frac{1}{N} \sum_{j=1}^N \log L(\hat\Lambda^{(-j)},\hat\Psi^{(-j)};\mathcal{X}_j),
\end{equation}
where $L$ is the LINFA likelihood function of the data $\mathcal{X}_j$ with the same form as Equation~\eqref{eq:linfalikelihood}, and $(\hat\Lambda^{(-j)},\hat\Psi^{(-j)})$ is the MLE of $(\Lambda,\Psi)$ based on all data except $\mathcal{X}_j$. The number of factors $q_{\rm CV}$ selected via N-CV is the minimizer of $\textsc{Risk}_{\rm NCV}(q)$.

Following the form of AIC for factor analysis \citep{akaike1987factor}, we define the LINFA AIC risk
\begin{equation}\label{eq:AIC}
    \textsc{Risk}_{\rm AIC}(q) ~=~ -2\log L(\hat\Lambda,\hat\Psi;\mathbf{X}^{(1)},\ldots,\mathbf{X}^{(K)})+2\kappa,
\end{equation}
where $L$ is the LINFA likelihood function in Equation~\eqref{eq:linfalikelihood}, and 
$\kappa$ is the number of free parameters (Equation~\eqref{eq:kappa}). The number of factors $q_{\rm AIC}$ selected via AIC is the minimizer of $\textsc{Risk}_{\rm AIC}(q)$. Finally, in Theorem~\ref{theo:BIC} in Appendix~\ref{app:modelsel}, we derive the LINFA BIC risk,
\begin{equation}\label{eq:BIC}
\textsc{Risk}_{\rm BIC}(q) ~=~ -2\log L(\hat\Lambda,\hat\Psi;\mathbf{X}^{(1)},\ldots,\mathbf{X}^{(K)})+\kappa\log n,
\end{equation} 
where $\kappa$ is the number of free parameters in Equation~\eqref{eq:kappa}.

In view of Theorem~\ref{lemma:rotlinkage}, it is wise to search for $q\le m^*$, where 
\begin{equation}\label{eq:m.star}
m^*=\min\left\{m_0,\lceil(d-1)/2\rceil-1\right\},
\end{equation}
and $m_0$ is the largest $m$ such that condition A1 holds.  
In the simulations in Section~\ref{sec:modelsel}, we observe that N-CV, AIC, and BIC adequately recover the true number of factors. However, BIC appears to be slightly more accurate than N-CV and AIC, so it is used in the data analysis in Section~\ref{sec:data}.

\section{Simulations}\label{sec:simulations}\vspace{-3mm}
In this section, we present the results of an extensive simulation study on the performance of LINFA. In Section~\ref{sec:simsettings}, we specify the simulation settings. In Section~\ref{sec:simuncertainty}, we verify the accuracy of the methods for uncertainty quantification proposed in Section~\ref{sec:uncertainty}. In Section~\ref{sec:modelsel}, we evaluate the performance of the model selection criteria proposed in Section~\ref{sec:modelselection}. In Section~\ref{sec:comptimes}, we assess the computational advantages of using the GVT algorithm in the LINFA MLE EM Algorithm~\ref{algo:mleEM}. Finally, in Section~\ref{sec:modelcomp}, we compare the performance of LINFA with other approaches in (a) estimating the full covariance matrix $\Sigma=\Lambda\Lambda^\T+\Psi$; (b) estimating the conditional dependence among the variables; (c) recovering $\Lambda\Lambda^\T$ and $\Psi$ individually; (d) predicting the latent factors; and (e) completing the data.

\subsection{Simulation settings}~\label{sec:simsettings}
We generate zero mean multivariate Gaussian data with a ground truth covariance matrix $\Sigma=\Lambda\Lambda^\T+\Psi$, where the diagonals of $\Psi$ were drawn without replacement from a set of $d$ evenly spaced values in the interval $[1/d,5]$, and $\Lambda$ was initially filled with values drawn without replacement from a set of $dq$ evenly spaced values in the interval $[-2,2]$, then rotated via Equation~\eqref{eq:rotMLE}, and finally had its signs flipped via Equation~\eqref{eq:signflip}. 
We set
\begin{equation}\label{eq:simvk}
V_k=\left\{1+\left\lfloor\tfrac{k-1}{K-1}(d-d_0)\right\rfloor, \ldots., d_0+\left\lceil\tfrac{k-1}{K-1}(d-d_0)\right\rceil\right\}, ~k=1,\ldots,K,
\end{equation} 
so that $|V_k|\approx d_0$, where $d_0\in(d/K,d)$ is taken to achieve the desired missingness proportion  
\begin{equation}\label{eq:eta}
\eta:=|O^c|/d^2,
\end{equation}
where $O:=\cup_{k=1}^KV_k\times V_k$ is the set of observed variable pairs. Finally, the sample size of each data set $\mathbf{X}^{(k)}$ is $n_k=\lceil n/K\rfloor$. It can be verified that, in all simulation settings, assumptions A1--A4 hold with $p_k=1/K$, for all $k=1,\ldots,K$, and $b\in(0,(5+dq4)^{-1}]$.

\begin{figure*}[t!]
    \centering 
\includegraphics[width=1\textwidth]{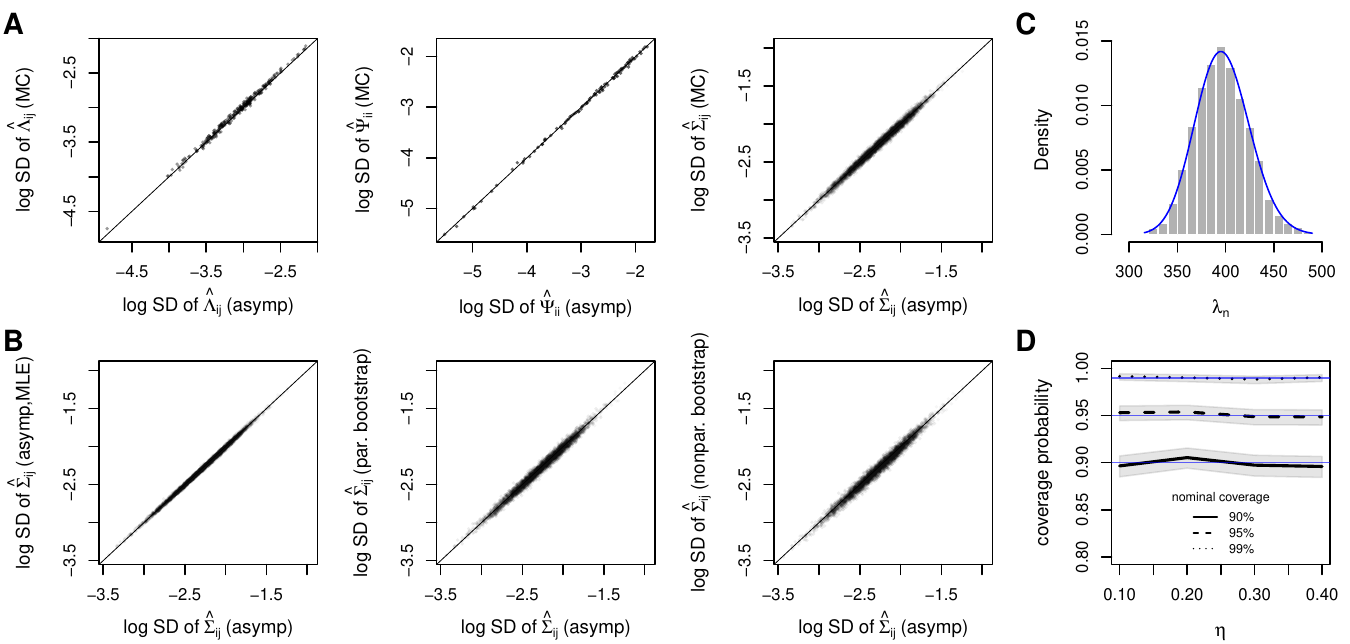}
\caption{\textbf{Uncertainty quantification} (settings: $d=100$, $q=3$, $\eta=0.2$, $K=3$, and $n=5000$).  \textbf{(A)} Log SD of the LINFA MLEs $\hat\Lambda_{ij}$, $\hat\Psi_{ii}$, and $\hat\Sigma_{ij}$ approximated via Monte Carlo (500 repeats) and plotted against their asymptotic approximations (Equations~\eqref{eq:asnorm}, \eqref{eq:acovsigma}). \textbf{(B)} Log SD of the LINFA MLE  $\hat\Sigma_{ij}$ estimated via MLE, nonparametric bootstrap, and parametric bootstrap plotted against the asymptotic approximations. 
\textbf{(C)} Histogram of the likelihood ratio statistic $\lambda_n$ (Equation~\eqref{eq:confreg}; 5000 repeats). The histogram approximates well the theoretical p.d.f.~of $\chi^2_\kappa$ (continuous curve), with degrees of freedom $\kappa=397$ as per Theorem~\ref{theo:confreg}. 
\textbf{(D)} Estimated coverage probability of the confidence region for $(\Lambda,\Psi)$ (Equation~\eqref{eq:confreg}) for three levels of nominal coverage $90\%,95\%,99\%$ and $0.1\le\eta\le 0.4$.
    }
    \label{fig:uncertainty}
\end{figure*}

\subsection{Uncertainty quantification}\label{sec:simuncertainty}
In this simulation, we assess the accuracy of the methods for uncertainty quantification proposed in Section~\ref{sec:uncertainty}, using the settings $d=100$, $q=3$, $\eta=0.2$, $K=3$, and $n=5000$. 
In Figure~\ref{fig:uncertainty}A, we compare the standard deviations of the LINFA MLE $\hat\Lambda_{ik}$, $\hat\Psi_{ii}$, and $\hat\Sigma_{ij}$ for all $i,j,k$, approximated via Monte Carlo (500 repeats), with those obtained asymptotically in Equations~\eqref{eq:acov} and \eqref{eq:acovsigma}. The two approximations are very close, supporting the accuracy of the asymptotic formulas. 
In Figure~\ref{fig:uncertainty}B, we compare the theoretical standard deviations of the LINFA MLE $\hat\Sigma_{ij}$, approximated via Equation~\eqref{eq:acovsigma}, with their MLEs and with the estimates obtained via parametric bootstrap (Algorithm~\ref{algo:parbootstrap}) and nonparametric bootstrap (Algorithm~\ref{algo:nonparbootstrap}). All estimators adequately recover the theoretical standard deviations. 

In Figure~\ref{fig:uncertainty}C, we show the histogram of the likelihood-ratio statistic $\lambda_n$ (Equation~\eqref{eq:confreg}) based on 5000 repeats. The histogram approximates the theoretical p.d.f.~of $\chi^2_\kappa$ (blue curve) well, with degrees of freedom $\kappa=397$ as per Theorem~\ref{theo:confreg}.
Finally, in Figure~\ref{fig:uncertainty}D, we display the estimated coverage probability (5000 repeats) of the confidence region in Equation~\eqref{eq:confreg} for three levels of nominal coverage $90\%,95\%,99\%$ and missingness proportion $0.1\le\eta\le 0.4$. The estimated coverage probability accurately reflects the nominal coverage.

\subsection{Model selection}\label{sec:modelsel}
\begin{figure*}[t!]
    \centering
\includegraphics[width=.7\textwidth]{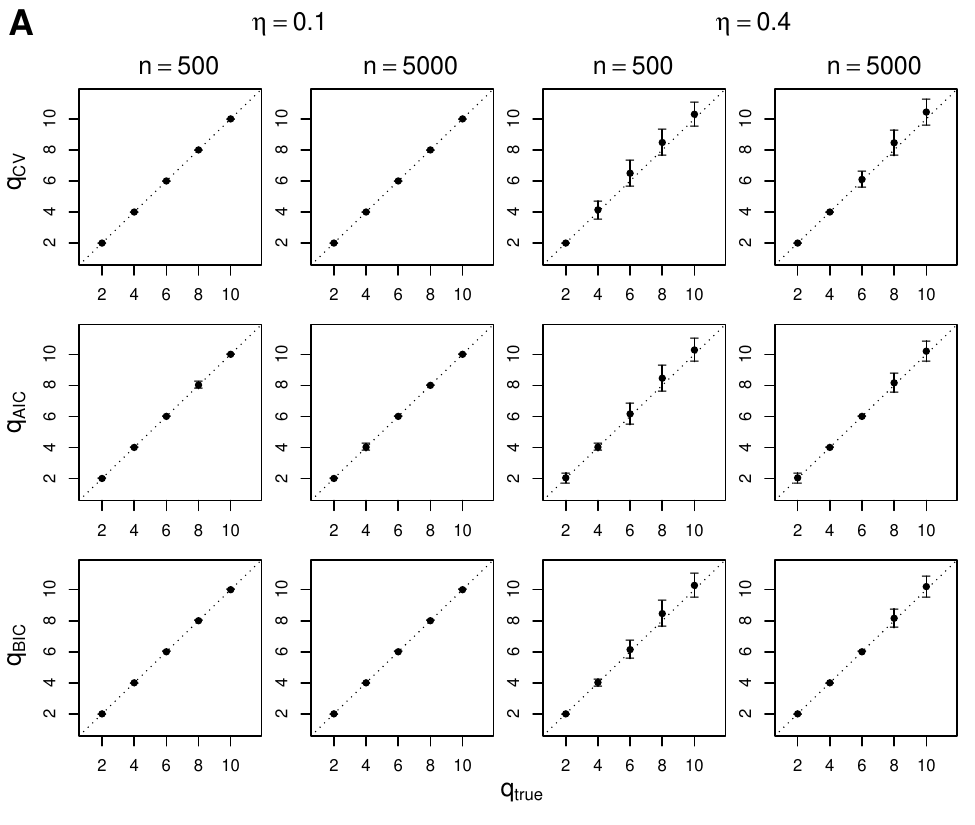}\includegraphics[width=.3\textwidth]{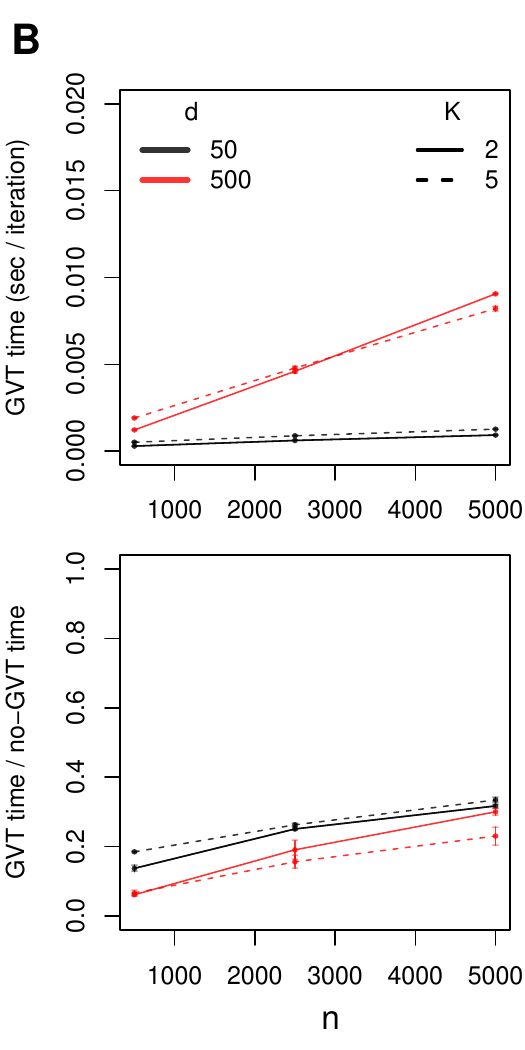}
    \caption{\textbf{(A)} Average ($\pm$ 2SD) selected number of factors $q_{\rm CV}$, $q_{\rm AIC}$, and $q_{\rm BIC}$ versus ground truth $q_{\rm true}$ (settings: $K=4$; $d=100$). 
    \textbf{(B)} Computational time of the LINFA Algorithm~\ref{algo:mleEM} (settings: $q=2$, $\eta=0.4$) using the GVT Algorithm~\ref{algo:gvt} in sec/iteration (top) and as proportion of the time required when optimizing coordinate-wise (bottom).
    }
    \label{fig:simmodelsel}
\end{figure*}

In Figure~\ref{fig:simmodelsel}A, we display the average selected number of factors $q_{\rm CV}$, $q_{\rm AIC}$, and $q_{\rm BIC}$, obtained by minimizing $\textsc{Risk}_{\rm 2CV}$ (Equation~\eqref{eq:cv}), $\textsc{Risk}_{\rm AIC}$ (Equation~\eqref{eq:AIC}), and  $\textsc{Risk}_{\rm BIC}$ (Equation~\eqref{eq:BIC}), respectively, versus the ground truth value $q_{\rm true}$. The three criteria appear to yield similar selections and adequately recover the true number of factors under various settings with $K=4$, $d=100$, $n=500,5000$, $q_{\rm true}=2,4,6,8,10$, and $\eta=0.1,0.4$. However, BIC appears to be slightly more accurate, so we use it in our data analysis in Section~\ref{sec:data}. 

~

~

\subsection{Computational times}\label{sec:comptimes}
We compare the computational efficiency of the LINFA MLE EM Algorithm~\ref{algo:mleEM}, implemented with the sets $W_1,\ldots, W_J$ produced by the GVT Algorithm~\ref{algo:gvt}, to the one implemented with the sets $W_1,\ldots, W_d$ defined as $W_i=i$, i.e., coordinate-wise, in the settings $q=2$, $K=2, 5$, $d=50,500$, $n=500,2500,5000$, and $\eta=0.4$. In the top panel of Figure~\ref{fig:simmodelsel}B, we display the average time per iteration required by the LINFA MLE algorithm, utilizing the GVT Algorithm~\ref{algo:gvt}. In the bottom panel, we divide these average times by those required when optimizing coordinate-wise. The GVT algorithm yields a substantially faster computation of the LINFA MLE, especially for larger $d$.

\subsection{Model comparisons}\label{sec:modelcomp}
\begin{figure*}[t!]
    \centering 
\includegraphics[width=1\textwidth]{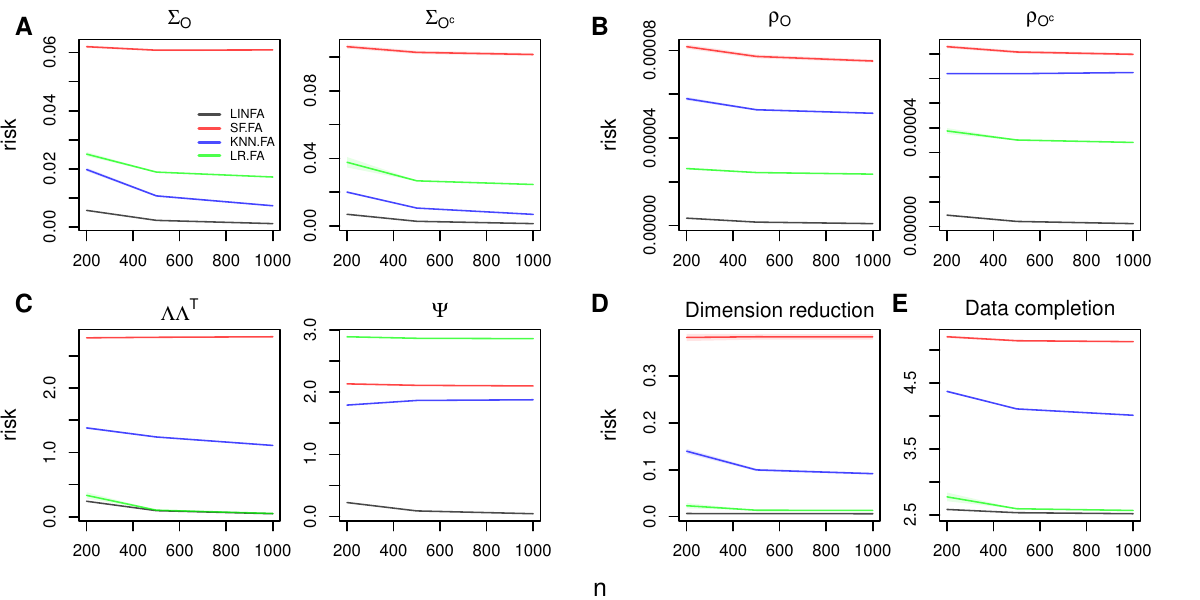}
    \caption{{\bf Model comparison} in the settings $d=200$, $q=2$, $K=4$, and $\eta=0.4$. \textbf{(A)}. Correlation estimation risk in $O$ and $O^c$. 
    \textbf{(B)} Partial correlation estimation risk in $O$ and $O^c$. 
    \textbf{(C)} Estimation risk for the components $\Lambda\Lambda^\T$ and $\Psi$. 
    \textbf{(D)} Dimension reduction risk. 
    \textbf{(E)} Data completion risk. 
    }
    \label{fig:simperformance}
\end{figure*}

We compare the performance of LINFA with three alternative approaches that fit traditional FA on data completed through various existing completion methods. Specifically: (a) {\it simple fill \& factor analysis} (\textbf{SF-FA}) fits FA on a completed $n\times d$ data matrix $\mathbf{X}_{\rm SF}$ obtained by filling any missing data point about the $j$-th variable with the average of all observed values of variable $j$; (b) {\it k nearest neighbors \& factor analysis} (\textbf{KNN-FA}) fits FA on a completed $n\times d$ data matrix $\mathbf{X}_{\rm KNN}$ obtained via KNN matrix completion \citep{vinci2024unsupervised}; and (c) {\it low-rank \& factor analysis} (\textbf{LR-FA}) fits FA on a completed data matrix $\mathbf{X}_{\rm LR}$ obtained via low-rank matrix completion \citep{mazumder2010spectral,vinci2024unsupervised}.  In Figure~\ref{fig:simperformance}, we summarize the simulation results in the settings $d=200$, $q=2$, $K=4$, $\eta=0.4$, and $200\le n\le 1000$. All results are presented with 95\% confidence intervals.

In Figure~\ref{fig:simperformance}A, we compare the methods' correlation estimation accuracy. For each method, we plot the average of the mean squared distance between the correlations $C_{ij}=\Sigma_{ij}(\Sigma_{ii}\Sigma_{jj})^{-1/2}$ and their estimates $\hat C_{ij}=\hat\Sigma_{ij}(\hat\Sigma_{ii}\hat\Sigma_{jj})^{-1/2}$ in the set $O$ of observed pairs of variables and in the set $O^c$ of unobserved variable pairs. LINFA outperforms all other methods.

In Figure~\ref{fig:simperformance}B, we present results analogous to Figure~\ref{fig:simperformance}A, but we compare the partial correlations $\rho_{ij}=-\Theta_{ij}(\Theta_{ii}\Theta_{jj})^{-1/2}$ with their estimates $\hat\rho_{ij}=-\hat\Theta_{ij}(\hat\Theta_{ii}\hat\Theta_{jj})^{-1/2}$. Even in this case, LINFA outperforms all other methods.

In Figure~\ref{fig:simperformance}C, we compare the accuracy of the estimation of the components $\Lambda\Lambda^\T$ and $\Psi$. For each method, we plot the average of the mean squared difference between the entries of $\Lambda\Lambda^\T$ and $\hat\Lambda\hat\Lambda^\T$, and between the diagonals of $\Psi$ and $\hat\Psi$. LINFA outperforms all other methods.  LR-FA recovers the low rank component $\Lambda\Lambda^\T$ well, but it recovers $\Psi$ poorly. 

In Figure~\ref{fig:simperformance}D, we compare the latent factor prediction accuracy of the methods. LINFA predicts the latent factors via Equation~\eqref{eq:factorpred}, while SF-FA, LR-FA, and KNN-FA use the equation $\hat Z_r=\hat\Lambda^\T\hat\Sigma^{-1}\tilde X_r$, where $\hat\Lambda$ and $\hat\Sigma$ are estimates from the FA model fitted on the completed data vectors $\tilde X_1,\ldots,\tilde X_n$ contained in $\mathbf{X}_{\rm SF}$, $\mathbf{X}_{\rm KNN}$, and $\mathbf{X}_{\rm LR}$, respectively. We assess the accuracy of factor prediction using the \textit{trace} $R^2$ metric of the regression of the estimated factors on the true ones \citep{stock2002forecasting, doz2012quasi, banbura2014maximum}, which is defined as $R^2(Z,\hat Z) = {\rm tr}(Z^\T\hat Z(\hat Z^\T\hat Z)^{-1}\hat Z^\T Z)/{\rm tr}(Z^\T Z)$, where $Z\in\mathbb{R}^{n\times q}$ is the matrix containing the $n$ generated factor vectors $Z_1,\ldots,Z_n$, and $\hat Z$ is the predicted counterpart. In the figure, we plot the average loss $1-R^2(Z,\hat Z)$. LINFA outperforms all other methods; LR-FA also appears predict the factors well.
 
Finally, in Figure~\ref{fig:simperformance}E, we compare the data completion accuracy of the methods. LINFA completes the data via Equation~\eqref{eq:datacompl}, while the other methods complete the data before fitting FA, as described at the beginning of Section~\ref{sec:simulations}. We calculate the average of the mean squared difference between the unobserved data values and their predictions. LINFA outperforms all other methods.

\section{Application to neuroscience}\label{sec:data}\vspace{-3mm}
Estimating the covariance matrix of neurons' activities is a fundamental step in the analysis of functional neural connectivity, the statistical dependence among neurons \citep{vinci2016separating,vinci2018adjusted,vinci2018adjustedcovariance}. The electrophysiological activity of thousands of neurons is now commonly recorded using cutting-edge technologies such as 2-photon calcium imaging \citep{stringer2019spontaneous,microns2021functional}. However, due to technological limitations, recording from large populations of neurons with fine temporal resolution is often impossible, and it is preferable to record subsets of neurons at once with a fine temporal resolution. This procedure may yield structurally incomplete data.

We apply LINFA to the analysis of publicly available neuroscience data  \citep{stringer2019spontaneous} consisting of calcium activity traces recorded from about 10,000 neurons in a 1mm $\times$ 1mm $\times $ 0.5mm volume of mouse visual cortex (70–385µm depth). The neuronal activities were simultaneously recorded \textit{in vivo} via 2-photon imaging of GCaMP6s with a 2.5 Hz scan rate \citep{pachitariu2017suite2p}. During the experiment, the animal was free to run on an air-floating ball in complete darkness for 105 minutes. For our analysis, we focus on the 725 neurons occupying the most superficial layer of the recorded brain. To mitigate time dependence and make the data closer to the normal distribution, the calcium activity traces are summed in groups of three consecutive time bins and then
square-rooted, yielding a total sample size of $n=5253$. To validate LINFA, we artificially obliterate portions of the data to generate various levels of structural missingness, and we compute the LINFA MLE in each scenario and compare it to the estimate obtained from the complete data.

\begin{figure*}[t!]
    \centering
\includegraphics[width=1\textwidth]{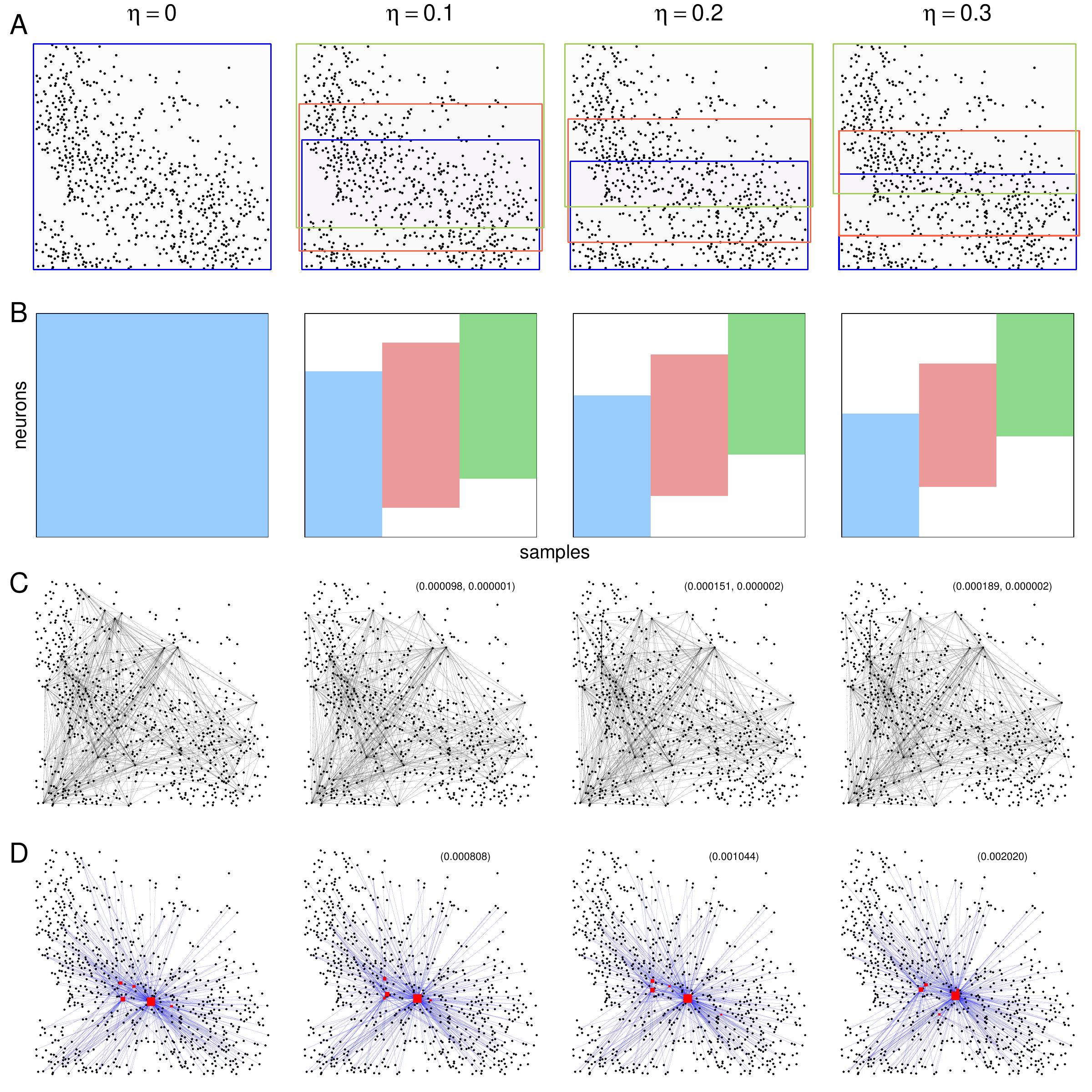}
    \caption{\textbf{(A)} Observed subsets of simultaneously recorded neurons inducing proportions of missingness $\eta=0,0.1,0.2,0.3$. \textbf{(B)} Observed data patterns. \textbf{(C)} Partial correlation graphs (top 400 edges). In parentheses: average square distance between correlations and partial correlations from the complete data case ($\eta=0$). \textbf{(D)} Factor graphs (top 400 edges) with $q=5$ factor nodes (red squares) with positions as per Equation~\eqref{eq:factorpositions}. In parentheses: average square distance between the magnitudes of $\gamma_{ij}$ from the complete data case ($\eta=0$).
}
\label{fig:data}
\end{figure*}

Figure~\ref{fig:data}A shows the physical position of the neurons on the 2-dim slice with four observation scenarios with different proportions of missingness $\eta=0,0.1,0.2,0.3$ (Equation~\eqref{eq:eta}).  Figure~\ref{fig:data}B shows the corresponding observed data patterns. 
We estimate LINFA in each scenario with a number of factors $q$ selected via BIC (Section~\ref{sec:modelselection}). For the cases $\eta=0,0.1,0.2,0.3$, BIC  selected $q=9,7,6,5$, respectively, which are values well below the maximum allowed to satisfy the linkage condition ($m^*=361,326,238$ for $\eta=0.1,0.2,0.3$, respectively; Equation~\eqref{eq:m.star}). However, to make the results easier to compare, we fit LINFA with the smallest selected number of factors, $q=5$, for all cases. 

In Figure~\ref{fig:data}C, we show partial correlation graphs (top 400 edges with the largest $|\hat\rho_{ij}|$, Equation~\eqref{eq:rho}). In parentheses, we indicate the average squared difference between estimated correlations and partial correlations from the complete data case ($\eta=0$). LINFA robustly recovers correlations and partial correlations for all levels of missingness. In Figure~\ref{fig:data}D, we show the estimated factor graphs (top 400 edges with the largest $|\hat\gamma_{ij}|$, Equation~\eqref{eq:gamma}), where the factor nodes (red squares) have positions computed as the average of the neurons coordinates with weights proportional to the magnitudes of the $\gamma_{ij}$'s. Specifically, if $Y\in\mathbb{R}^{d\times 2}$ contains the 2-dim coordinates of the $d$ neurons, then the position of the $j$-th factor is given by 
\begin{equation}\label{eq:factorpositions}
  y_j ~=~  \frac{\sum_{i=1}^d |\gamma_{ij}|Y_i}{\sum_{i=1}^d|\gamma_{ij}|},
\end{equation}
where $Y_i$ is the $i$-th row of $Y$. 
It is interesting to note that the weighted positions of the most important factors (factors with the largest $\sum_{i=1}^d |\gamma_{ij}|$) are approximately the same across different levels of missingness. In the figure, we indicate the average squared difference between the estimated magnitudes of $\gamma_{ij}$ from the complete data case ($\eta=0$) in parentheses. LINFA robustly recovers the conditional correlations $\gamma_{ij}$'s for all levels of missingness.

\section{Conclusions}\label{sec:discussions}\vspace{-3mm}
We proposed the Linked Factor Analysis model (LINFA), a unified statistical framework designed to address challenges arising from structurally incomplete data. This type of missingness occurs when the data are not fully observed, and several pairs of variables may lack joint observations. LINFA simultaneously integrates four key tasks: covariance matrix completion, dimensionality reduction, data completion, and dependence structure recovery. 

We established the conditions for the uniqueness, consistency, and asymptotic normality of the LINFA MLE of the parameters $(\Lambda,\Psi)$ and the covariance matrix $\Sigma=\Lambda\Lambda^\T+\Psi$. One of these conditions is $q$-linkage, which prescribes the minimum overlap among the observed variable sets $V_1,\ldots,V_K$ required to preclude the arbitrary rotation of any subset of the loading matrix estimate. We further derived confidence regions and hypothesis tests based on the asymptotic likelihood ratio and proposed bootstrap methods.

To compute the LINFA MLE, we proposed an efficient Expectation-Maximization algorithm in which all updating equations are in closed form, and the maximization step is significantly accelerated using the Group Vertex Tessellation (GVT) algorithm. The GVT algorithm allows us to identify a minimal partition $W_1,\ldots,W_J$ of the vertex set $V=\{1,\ldots,d\}$, where all nodes in $W_k$ are observed simultaneously on exactly the same samples across the data sets $\mathbf{X}^{(1)},\ldots,\mathbf{X}^{(K)}$, allowing for the implementation of the M-step with closed form expressions and the smallest possible number of operations. We further designed various approaches for the empirical selection of the number of factors by adapting N-fold cross-validation (N-CV), the Akaike Information Criterion (AIC), and the Bayesian Information Criterion (BIC) to our incomplete data settings. 

We demonstrated the performance of LINFA in an extensive simulation study. The uncertainty quantification techniques based on the asymptotic normality of the LINFA MLE and bootstrap procedures accurately assess the variability of the estimators. The proposed multidimensional confidence region provides an actual coverage probability close to the nominal one. The developed LINFA N-CV, AIC, and BIC accurately select the true number of factors. The LINFA algorithm is efficient, especially with the acceleration provided by the GVT algorithm. LINFA outperforms other existing methods in covariance matrix completion, partial correlation estimation, dimensionality reduction, and data completion.

Finally, we applied LINFA to the analysis of neuronal data recorded \textit{in vivo} from a mouse's visual cortex via 2-photon calcium imaging technology. LINFA accurately recovered the  functional connectivity structure of hundreds of neurons from incomplete data with a precision comparable to standard FA applied to the full data scenario. 

 We expect LINFA to become a common tool for the multivariate analysis of data from disparate scientific fields where structural missingness may arise, such as neuroscience, genomics, psychology, medicine, and astronomy. 
 While in this article we approached the LINFA estimation task via MLE, other approaches, such as penalized MLE and Bayesian modeling, may be considered. We intend to investigate these approaches in the future.

\section*{Code and data availability}\vspace{-3mm} 
The R-package {\tt linfa} is available on GitHub at \url{https://github.com/gvincistat}. 
The neuroscience data set \citep{stringer2019spontaneous} is publicly available at the following URL: \url{https://figshare.com/ndownloader/files/11152646}.

\bibliographystyle{apalike}
\bibliography{bibliography.bib}

 \newpage

\begin{center}
    {\Large \bf Appendix}
\end{center}
\appendix

\section{Proofs of Main Results}\label{app:proofs}
In this appendix, we prove all results presented in Section~\ref{sec:mle}.

\subsection{Linkage}\label{app:linkage}

\begin{proof}[Proof of Lemma~\ref{lemma:rotlinkage}]
First of all, note that the equation
\begin{equation}\label{eq:mleequivx}
\Sigma^{(k)}=\tilde\Sigma^{(k)},~\forall k=1,\ldots,K
\end{equation}
holds only if $\Psi=\tilde\Psi$, because we need 
\begin{equation}
{\rm diag}(\Psi^{(k)})-{\rm diag}(\tilde\Psi^{(k)})={\rm diag}(\tilde\Lambda^{(k)}\tilde\Lambda^{(k)\T})-{\rm diag}(\Lambda^{(k)}\Lambda^{(k)\T})=0
\end{equation}
otherwise the off-diagonals of $\Lambda^{(k)}\Lambda^{(k)\T}$ and $\tilde\Lambda^{(k)}\tilde\Lambda^{(k)\T}$ would be forced to be different, making Equation~\eqref{eq:mleequivx} impossible to hold. Thus, Equation~\eqref{eq:mleequivx} holds if and only if
\begin{equation}
\Psi~=~\tilde\Psi
\end{equation}
and
\begin{equation}\label{eq:LLunique}
\Lambda^{(k)}\Lambda^{(k)\T}~=~\tilde\Lambda^{(k)}\tilde\Lambda^{(k)\T}, ~\forall k=1,\ldots,K
\end{equation} 
For any $j,k$ such that $|V_k\cap V_j|\ge q$, it is impossible to alter $\tilde\Lambda_{V_k}$ while keeping $\tilde\Lambda_{V_j}$ fixed and Equation~\eqref{eq:LLunique} hold. Indeed, consider the alternative solution $(\Lambda,\Psi)$ where $\Lambda_{V_k\setminus V_j}=\tilde\Lambda_{V_k\setminus V_j}R$, $\Lambda_{(V_k\setminus V_j)^c}=\tilde\Lambda_{(V_k\setminus V_j)^c}$, $\Psi=\tilde\Psi$, and $R$ is a $q\times q$ rotation matrix, which has the property
\begin{equation}\label{eq:RR}
RR^\T=I_q
\end{equation}
Then, Equation~\eqref{eq:LLunique} holds if and only if 
\begin{equation}\label{eq:Lkj}
\Lambda_{V_k\setminus V_j}\Lambda_{V_k\cap  V_j}^\T :=\tilde\Lambda_{V_k\setminus V_j}R \tilde\Lambda_{V_k\cap  V_j}^\T = \tilde\Lambda_{V_k\setminus V_j} \tilde\Lambda_{V_k\cap  V_j}^\T,
\end{equation}
where $\Lambda_{V_k\cap  V_j} =\tilde \Lambda_{V_k\cap  V_j} $ because $V_k\cap V_j\subset (V_k\setminus V_j)^c$. 
Since $|V_k\cap V_j|\ge q$, by post-multiplying the left-hand and the right-hand sides of Equation~\eqref{eq:Lkj} by the matrix
\begin{equation}
\tilde\Lambda_{V_k\cap  V_j}\left(\tilde\Lambda_{V_k\cap  V_j}^\T\tilde\Lambda_{V_k\cap  V_j}\right)^{-1},
\end{equation}
we obtain the condition
\begin{equation}
\Lambda_{V_k\setminus V_j} = \tilde\Lambda_{V_k\setminus V_j}  
\end{equation}
Therefore, $\tilde\Lambda_{V_k\setminus V_j}$, and thereby $\tilde\Lambda_{V_k}$, cannot be altered given $\tilde\Lambda_{V_j}$.  
If the sets $V_1,\ldots,V_K$ are $q$-linked (Definition~\eqref{def:linkage}), for every pair $(V_i,V_j)$, there exists at least one sequence $V_{s_1},\ldots,V_{s_M}$, with distinct indices $1\le s_1,\ldots,s_M\le K$ and $s_1=i$, $s_M=j$, such that $|V_{s_h}\cap V_{s_{h+1}}|\ge q$, for all $h=1,\ldots,M-1$, so that  
$\tilde\Lambda_{V_{h+1}}$ cannot be altered given $\tilde\Lambda_{V_h}$. Consequently, $\tilde\Lambda_{V_j}$ cannot be altered given $\tilde\Lambda_{V_i}$, and similarly $\tilde\Lambda_{V_i}$ cannot be altered given $\tilde\Lambda_{V_j}$. This holds for all $\{i,j\}\subseteq\{1,\ldots,K\}$. This proves the sufficiency of the condition of the theorem. To prove its necessity, note that if the sets $V_1,\ldots,V_K$ are \textit{not} $q$-linked, then there exists at least one pair $V_i,V_j$ that is completely disconnected in the linkage graph $G^{(q)}$, or that is connected only by paths where at least two consecutive components $h,l$ have $|V_h\cap V_l|<q$, for which an arbitrary rotation $R$ would be allowed. 
\end{proof}

\begin{proof}[Proof of Theorem~\ref{theo:mlelinkage}]
The LINFA likelihood function in Equation~\eqref{eq:linfalikelihood} depends on $(\Lambda,\Psi)$ only through the components $\Sigma^{(1)},\ldots,\Sigma^{(K)}$, where $\Sigma^{(k)}=\Lambda^{(k)}\Lambda^{(k)\T}+\Psi^{(k)}$. Thus, if $(\tilde\Lambda,\tilde\Psi)$ is an MLE, any solution $(\breve\Lambda,\breve\Psi)$ that yields 
\begin{equation}\label{eq:sss}
\breve\Sigma^{(k)}=\tilde\Sigma^{(k)},~\forall k=1,\ldots,K
\end{equation}
or, equivalently, $\breve\Sigma_O=\tilde\Sigma_O$ with $O=\cup_{k=1}^KV_k\times V_k$, is also an MLE. By Lemma~\ref{lemma:rotlinkage}, we have that Equation~\eqref{eq:sss} implies $\breve\Lambda=\tilde\Lambda R$ and $\breve\Psi=\tilde\Psi$ for some rotation matrix $R$ if and only if the sets $V_1,\ldots,V_K$ are $q$-linked, i.e., if and only if the LINFA likelihood function is $q$-linked.
\end{proof}

\subsection{EM Algorithm}\label{app:EM}
In this appendix, we provide the proof of the optimality of the GVT Algorithm~\ref{algo:gvt} (Theorem~\ref{theo:gvt}), the derivation of the closed formulas of the EM Algorithm~\ref{algo:mleEM} (Theorem~\ref{theo:MLEalgoUpEq}), as well as the proof of its algorithmic convergence (Theorem~\ref{theo:MLEalgoconv}).

\begin{proof}[Proof of Theorem~\ref{theo:gvt}] 
Recall $I\in\{0,1\}^{d\times K}$, where $I_{ik}=1$ if $i\in V_k$, and $I_{ik}=0$ otherwise, and  $\Delta\in\mathbb{R}_+^{d\times d}$, where $\Delta_{ij}=\Vert I_{i.}-I_{j.} \Vert_1 $. Moreover, let $\mathcal{W}=\big\{\{j\in V:\Delta_{ij}=0\}: i\in V\big\}$.

(i) For any $i\in V$, we have $\{j\in V:\Delta_{ij}=0\}\supseteq \{i\}$ since $\Delta_{ii}=0$ by definition. Thus, $\cup_{j=1}^JW_j =\cup_{W\in\mathcal{W}}W=\cup_{i\in V}\{j\in V:\Delta_{ij}=0\}\supseteq \cup_{i=1}^d\{i\}=V$. Moreover, $W_j\subseteq V$ for any $j$, so $\cup_{j=1}^J W_j\subseteq V$. Thus, $\cup_{j=1}^JW_j=V$. We now show that $W_1,\ldots, W_J$ are pairwise disjoint. For any $1\le i<j\le J$ we have $W_i\neq W_j$ by fundamental set theory (the set $\mathcal{W}$ contains no duplicates). Suppose $W_i\cap W_j\neq \emptyset$, for some $1\le i<j\le J$. Then, there must exist at least one $l\in V$ such that $l\in W_i$ and $l\in W_j$, so that $\Delta_{sl}=0$ and $\Delta_{tl}=0$, for all $s\in W_i$ and for all $t\in W_j$. But this would imply $W_i=W_j$, which is a contradiction. Thus, $W_1,\ldots,W_J$ are pairwise disjoint. Therefore, $W_1,\ldots,W_J$ is a partition of $V$.

(ii) If $W_1,\ldots, W_J$ is a partition of $V=\cup_{k=1}^K V_k$, the constraint in Equation~\eqref{eq:gvteq} holds if and only if, for every $j=1,\ldots,J$, the rectangle $W_j\times \mathcal{K}_{W_j}$ contains all pairs $(i,k)$ such that $I_{ik}=1$ with $i\in W_j$. Thus, we need $ \mathcal{K}_{W_j}=\mathcal{K}_{\{i\}}$, for all $i\in W_j$. Indeed, the partition $W_1,\ldots,W_J$, where $W_i=\{i\}$ and $J=d$, satisfies Equation~\eqref{eq:gvteq}. To have smallest $J$, we simply need to define $W_1,\ldots,W_J$ as sets of nodes where $\mathcal{K}_{W_l}\neq \mathcal{K}_{W_h}$, for all $h\neq l$, and $\mathcal{K}_{\{i\}}=\mathcal{K}_{W_j}$ for all $i\in W_j$, and all $j=1,\ldots,J$. That is, any pair of nodes $i$ and $l$ must be in the same set $W_j$ if and only if $\mathcal{K}_{\{i\}}=\mathcal{K}_{\{l\}}$.  The set $W_1,\ldots, W_J$ produced by Algorithm~\ref{algo:gvt} satisfy this property.
\end{proof}

\begin{theorem}[\sc Updating equations of the EM Algorithm~\ref{algo:mleEM}]\label{theo:MLEalgoUpEq}
The updating equations for the LINFA EM algorithm are
\begin{equation}\label{eq:upLambApp}
\Lambda_{{W_j},t+1} = \left(\sum_{k\in\mathcal{K}_{W_j}}\mathbf{X}^{(k)\T}_{W_j} M_t^{(k)}\right)\left(\sum_{k\in\mathcal{K}_{W_j}}S^{(k)}_t \right)^{-1}
\end{equation}
and
\begin{equation}\label{eq:upPsiApp}
\Psi_{{W_j}{W_j},t+1} = {\rm Diag}\left\{\sum\limits_{k\in\mathcal{K}_{W_j}}\left(\mathbf{X}_{W_j}^{(k)\T}\mathbf{X}^{(k)}_{W_j} -\Lambda_{{W_j},t+1}S^{(k)}_t\Lambda_{{W_j},t+1}^\T \right)\right\}\left(\sum\limits_{k\in\mathcal{K}_{W_j}}n_k\right)^{-1},
\end{equation}
where 
\begin{eqnarray}
S_t^{(k)} &=& n_k(I_q-\Gamma^{(k)\T}_t\Lambda^{(k)}_t)+M^{(k)\T}_t M_t^{(k)},\\
M_t^{(k)} &=& \mathbf{X}^{(k)}\Gamma^{(k)}_t,\\
\Gamma^{(k)}_t &=& A^{(k)}_t(I_q-(I_q+B^{(k)}_t)^{-1}B^{(k)}_t)\\
A^{(k)}_t &=& \Psi^{(k)-1}_t\Lambda^{(k)}_t\\
B^{(k)}_t &=& A_t^\T\Lambda^{(k)}_t
\end{eqnarray}

\begin{proof}
The complete log-likelihood function is
\begin{eqnarray*}
&& \ell(\Lambda,\Psi; \{\mathbf{Z}^{(k)}\}_{k=1}^K, \{\mathbf{X}^{(k)}\}_{k=1}^K) = \sum_{k=1}^K \log L_k(\Lambda^{(k)},\Psi^{(k)}; \mathbf{Z}^{(k)}, \mathbf{X}^{(k)})\\
&& = \sum_{k=1}^K  \left\{-\frac{n_k}{2}\log\det\Psi^{(k)}-\frac{1}{2}{\rm tr}\left((\mathbf{X}^{(k)}-\mathbf{Z}^{(k)}\Lambda^{(k)\T})\Psi^{(k)-1}(\mathbf{X}^{(k)}-\mathbf{Z}^{(k)}\Lambda^{(k)\T})^\T\right)\right\}+C
\\
&& = -\frac{1}{2}\sum_{k=1}^K  \left\{n_k\log\det\Psi^{(k)}+{\rm tr}\left(\mathbf{X}^{(k)\T}\mathbf{X}^{(k)}\Psi^{(k)-1}\right)+{\rm tr}\left(\mathbf{Z}^{(k)\T}\mathbf{Z}^{(k)}\Lambda^{(k)\T}\Psi^{(k)-1}\Lambda^{(k)}
\right)\right.\\
& & ~~~~~~~~~- \left. 2{\rm tr}\left(\mathbf{Z}^{(k)}\Lambda^{(k)\T}\Psi^{(k)-1}\mathbf{X}^{(k)\T} \right)\right\}+C\nonumber
\end{eqnarray*}
where $C$ is a term that does not involve the parameters of interest $(\Lambda,\Psi)$. Then, the expectation of the complete log-likelihood function  given the data $\{\mathbf{X}^{(k)}\}_{k=1}^K$ and current estimates $\Lambda_t$, $\Psi_t$, is
\begin{eqnarray*}
Q_t(\Lambda,\Psi) &=& \E_t[\ell(\Lambda,\Psi; \{\mathbf{Z}^{(k)}\}_{k=1}^K, \{\mathbf{X}^{(k)}\}_{k=1}^K)\mid \{\mathbf{X}^{(k)}\}_{k=1}^K]\\
& = &   
-\frac{1}{2}\sum_{k=1}^K   \Big\{n_k\log\det \Psi^{(k)}+{\rm tr}\left(\mathbf{X}^{(k)\T}\mathbf{X}^{(k)}\Psi^{(k)-1} \right)  \\ 
& &~~~~~~   +{\rm tr}\left(S_t^{(k)}\Lambda^{(k)\T}\Psi^{(k)-1}\Lambda^{(k)}\right)
-2{\rm tr}\left(M_t^{(k)}\Lambda^{(k)\T}\Psi^{(k)-1}\mathbf{X}^{(k)\T} \right)\Big\}+{\rm const}
\end{eqnarray*}
where $\E_t[*]$ denotes expectation assuming the distribution of $\mathbf{X}^{(1)},\ldots,\mathbf{X}^{(K)}$ has parameters equal to $\Lambda_t,\Psi_t$, and by applying Lemma~\ref{lemma:condexpEM} and Woodbury Matrix Identity,
\begin{eqnarray*}
M_t^{(k)} &:=& \mathbb{E}_t[\mathbf{Z}^{(k)}\mid \mathbf{X}^{(k)}] ~=~ \mathbf{X}^{(k)}\Gamma^{(k)}_t,\\
S_t^{(k)} &:=& \mathbb{E}_t[\mathbf{Z}^{(k)\T}\mathbf{Z}^{(k)}\mid \mathbf{X}^{(k)}] ~=~ n_k(I_q-\Gamma^{(k)\T}_t\Lambda^{(k)}_t)+M^{(k)\T}_t M_t^{(k)},\\
\Gamma^{(k)}_t &:=&(\Lambda_t^{(k)}\Lambda_t^{(k)\T}+\Psi_t^{(k)})^{-1}\Lambda_t^{(k)} 
~=~ A^{(k)}_t(I_q-(I_q+B^{(k)}_t)^{-1}B^{(k)}_t),\\
A^{(k)}_t&:=&\Psi^{(k)-1}_t\Lambda^{(k)}_t,\\
B^{(k)}_t &:=& \Lambda^{(k)\T}_t\Psi^{(k)-1}_t\Lambda^{(k)}_t=
A_t^\T\Lambda^{(k)}_t
\end{eqnarray*}
Next, let $\mathcal{K}_{W_j}=\{k:W_j\subseteq V_k\}$ be the set of indices of the data sets where all nodes in $W_j$ were fully observed. The gradient of $Q_t$ with respect to the portion $\Lambda_{W_j}=[\Lambda_{il}]_{i\in W_j,l\in\{1,\ldots,q\}}$ is
\begin{eqnarray*}
\frac{\partial Q_t}{\partial\Lambda_{W_j}} &=& -\frac{1}{2}\sum_{k\in\mathcal{K}_{W_j}}\Big\{2[\Psi^{(k)-1}]_{{W_j}{W_j}}\Lambda_{W_j}S^{(k)}_t  -2[\Psi^{(k)-1}]_{{W_j}{W_j}}\mathbf{X}^{(k)\T}_{W_j}M_t^{(k)}\Big\}\\
&=& \sum_{k\in\mathcal{K}_{W_j}}\Big\{-\Psi^{-1}_{{W_j}{W_j}}\Lambda_{W_j}S^{(k)}_t 
+\Psi^{-1}_{{W_j}{W_j}}\mathbf{X}^{(k)\T}_{W_j}M_t^{(k)}\Big\}\\
&=& \Psi^{-1}_{{W_j}{W_j}}\left(-\Lambda_{W_j}\sum_{k\in\mathcal{K}_{W_j}}S^{(k)}_t 
+\sum_{k\in\mathcal{K}_{W_j}}\mathbf{X}^{(k)\T}_{W_j}M_t^{(k)}\right).
\end{eqnarray*} 

By setting this gradient to zero we obtain the updating equation
\begin{equation}\label{eq:upLambAPP}
\Lambda_{{W_j},t+1} = \left(\sum_{k\in\mathcal{K}_{W_j}}\mathbf{X}^{(k)\T}_{W_j} M_t^{(k)}\right)\left(\sum_{k\in\mathcal{K}_{W_j}}S^{(k)}_t \right)^{-1},
\end{equation}
where $\sum_{k\in\mathcal{K}_{W_j}}S^{(k)}_t$ is guaranteed to be positive definite, hence invertible, because for every $k$, 
\begin{equation}
S_t^{(k)} ~=~  n_k(I_q-\Lambda_t^{(k)\T}(\Lambda_t^{(k)}\Lambda_t^{(k)\T}+\Psi_t^{(k)})^{-1}\Lambda^{(k)}_t)+M^{(k)\T}_t M_t^{(k)}
\end{equation}
is positive definite as the first component is positive definite by  Lemma~\ref{lemma:posdefSt} in Appendix~\ref{app:lemmas}. Applying Equation~\eqref{eq:upLambAPP} for all $j=1,\ldots,J$ yields the updated $\Lambda_{t+1}$. 

Next, let $\Omega=\Psi^{-1}$. The gradient of $Q_t$ with respect to the portion $\Omega_{W_jW_j}$ is
\begin{eqnarray*}
\frac{\partial Q_t}{\partial \Omega_{{W_j}{W_j}}} &=& \frac{1}{2}\sum_{k\in\mathcal{K}_{W_j}}\left(n_k\Omega_{{W_j}{W_j}}^{-1} - \mathbf{X}^{(k)\T}_{W_j}\mathbf{X}^{(k)}_{W_j} -\Lambda_{{W_j}}S_t^{(k)}\Lambda^\T_{{W_j}} +2\mathbf{X}^{(k)\T}_{W_j}M_t^{(k)}\Lambda^{T}_{{W_j}}\right)
\end{eqnarray*}
By setting this gradient to zero and then plugging in $\Lambda_{W_j,t+1}$ in place of $\Lambda_{W_j}$ (Equation~\eqref{eq:upLambAPP}), we obtain the updating equation
\begin{equation}\label{eq:upPsiAPP}
\Psi_{{W_j}{W_j},t+1} = {\rm Diag}\left\{\sum\limits_{k\in\mathcal{K}_{W_j}}\left(\mathbf{X}_{W_j}^{(k)\T}\mathbf{X}^{(k)}_{W_j} -\Lambda_{{W_j},t+1}S^{(k)}_t\Lambda_{{W_j},t+1}^\T \right)\right\}\left(\sum\limits_{k\in\mathcal{K}_{W_j}}n_k\right)^{-1},
\end{equation}
where ${\rm Diag}(A)$ denotes the diagonal matrix with the same diagonal entries as the square matrix $A$ (i.e., ${\rm Diag}(A) = A\odot I$, where $I$ is the identity matrix and $\odot$ denotes the Hadamard product). Applying Equation~\eqref{eq:upPsiAPP} for all $j=1,\ldots,J$ yields the updated $\Psi_{t+1}$. The solution $(\Lambda_{t+1},\Psi_{t+1})$ is a maximum point because the Hessian matrix of $Q_t$ evaluated at $(\Lambda_{t+1},\Psi_{t+1})$ is negative definite, with negative definite diagonal blocks (for $j=1,\ldots,J$)
\begin{eqnarray*}
\frac{\partial^2 Q_t}{\partial\lambda_{W_j}\partial\lambda_{W_j}^\T} &=& -\sum_{k\in\mathcal{K}_{W_j}}S^{(k)}_t\otimes \Omega_{{W_j}{W_j}} \prec 0\\
\frac{\partial^2 Q_t}{\partial\omega_{W_j}\partial\omega_{W_j}^\T} &=& -\Omega_{W_jW_j}^{-1}\otimes \Omega_{{W_j}{W_j}}^{-1}\prec 0
\end{eqnarray*}
and null off-diagonal blocks (for $j\neq k$)
\begin{eqnarray*}
\frac{\partial^2 Q_t}{\partial\lambda_{W_j}\partial\omega_{W_j}^\T} &=& \left(-\Lambda_{W_j}\sum_{k\in\mathcal{K}_{W_j}}S^{(k)}_t 
+\sum_{k\in\mathcal{K}_{W_j}}\mathbf{X}^{(k)\T}_{W_j}M_t^{(k)}\right)^\T\otimes I_{|W_j|} ~=~ 0
\\
\frac{\partial^2 Q_t}{\partial\lambda_{W_j}\partial\lambda_{W_k}^\T} &=& 0
\\
\frac{\partial^2 Q_t}{\partial\lambda_{W_j}\partial\omega_{W_k}^\T} &=& 0
\\
\frac{\partial^2 Q_t}{\partial\omega_{W_j}\partial\omega_{W_k}^\T} &=& 0
\end{eqnarray*}
where $\lambda_{W_j}={\rm vec}(\Lambda_{W_j})$, $\omega_{W_j}={\rm vec}(\Omega_{W_jW_j})$, ${\rm vec}(A)$ is the vectorization of the matrix $A$, $\otimes$ denotes the Kronecker product, and $A\prec 0$ means that $A$ is a negative definite matrix.
\end{proof}
\end{theorem}

\begin{proof}[Proof of Theorem~\ref{theo:MLEalgoconv}]
The log-likelihood function based on Equation~\eqref{eq:linfalikelihood} is 
\begin{eqnarray*}
\ell\left(\Lambda,\Psi; \{\mathbf{X}^{(k)}\}_{k=1}^K\right) &=& \sum_{k=1}^K \log L_k(\Lambda^{(k)},\Psi^{(k)};\mathbf{X}^{(k)})\\ &=&
-\frac{1}{2}\sum_{k=1}^K n_k\left\{\log\det\Sigma^{(k)} + {\rm tr}\left(\hat\Sigma^{(k)}\Sigma^{(k)-1}\right)\right\}+C,
\end{eqnarray*}
where 
\begin{equation}\label{eq:sigmak}
\Sigma^{(k)}=\Lambda^{(k)}\Lambda^{(k)\T}+\Psi^{(k)},
\end{equation}
and
\begin{equation}\label{eq:sigmahatk}
\hat\Sigma^{(k)} = n_k^{-1} \mathbf{X}^{(k)\T}\mathbf{X}^{(k)},
\end{equation}
and $C$ is a constant terms that does not involve $\Lambda$ or $\Psi$. For any set $W=W_1,\ldots,W_J$ produced by the GVT Algorithm~\ref{algo:gvt}, the gradients with respect to $\Lambda_W$ and $\Psi_{W}$ are
\begin{eqnarray}\label{eq:gradLambda}
\frac{\partial\ell}{\partial \Lambda_W} &=& -\sum_{k\in\mathcal{K}_W}n_kU_{W.}^{(k)}\Lambda^{(k)}
\end{eqnarray}
and 
\begin{eqnarray}\label{eq:gradPsi}
\frac{\partial \ell}{\partial \Psi_{W}} &=& -\frac{1}{2}\sum_{k\in\mathcal{K}_W}n_k{\rm Diag}\left(U^{(k)}_{WW}\right)
\end{eqnarray}
where 
\begin{eqnarray}
U^{(k)} &=& \Sigma^{(k)-1}  - \Sigma^{(k)-1}\hat\Sigma^{(k)}\Sigma^{(k)-1}\\
&=& \Sigma^{(k)-1}(\Sigma^{(k)}-\hat\Sigma^{(k)})\Sigma^{(k)-1},
\end{eqnarray}
and the set 
\begin{equation}
\mathcal{K}_W = \{k:W\subseteq V_k \}
\end{equation}
indexes the datasets with observations about the node set $W$. Let $(\bar\Lambda,\bar\Psi)$, and corresponding $\bar\Sigma= \bar\Lambda\bar\Lambda^\T+\bar\Psi$, be a converging solution of the EM Algorithm~\ref{algo:mleEM} given fixed initial values for $\Lambda$ and $\Psi$. We are going to show that $(\bar\Lambda,\bar\Psi)$ satisfies the first order conditions $\frac{\partial\ell}{\partial\Lambda}=0$ and $\frac{\partial\ell}{\partial\Psi}=0$.

By exploiting Equations~\eqref{eq:Mt}--\eqref{eq:Gt}, we obtain
\begin{eqnarray*}
S_t^{(k)} &=& n_k(I_q-\Gamma^{(k)\T}_t\Lambda^{(k)}_t)+\Gamma_t^{(k)\T}\mathbf{X}^{(k)\T}\mathbf{X}^{(k)}\Gamma_t^{(k)}\\
&=& 
n_k(I_q-\Lambda^{(k)\T}_t\Sigma^{(k)-1}_t\Lambda^{(k)}_t)+n_k\Lambda_t^{(k)\T}\Sigma^{(k)-1}_t\hat\Sigma^{(k)}\Sigma^{(k)-1}_t\Lambda_t^{(k)}\\
&=& n_k\left\{I_q-\Lambda^{(k)\T}_t\Sigma^{(k)-1}_t(\Sigma^{(k)}_t-\hat\Sigma^{(k)})\Sigma^{(k)-1}_t \Lambda^{(k)}_t \right\}
\end{eqnarray*}
and define the $q\times q$ matrix
\begin{eqnarray}
   \bar S^{(k)} &=& \lim_{t\to\infty} S^{(k)}_t \\
   &=& n_k\left\{I_q-\bar\Lambda^{(k)\T}\bar\Sigma^{(k)-1}(\bar\Sigma^{(k)}-\hat\Sigma^{(k)})\bar\Sigma^{(k)-1}\bar\Lambda^{(k)} \right\}.\\
   &=& n_k\left\{I_q-\bar\Lambda^{(k)\T}\bar U^{(k)}\bar\Lambda^{(k)} \right\}.\label{eq:SBAR}
\end{eqnarray}
where 
\begin{equation}\label{eq:ubar}
\bar U^{(k)} ~=~ \bar\Sigma^{(k)-1}(\bar\Sigma^{(k)}-\hat\Sigma^{(k)})\bar\Sigma^{(k)-1}
\end{equation}
Thus, as $t\to\infty$, the updating equation of $\Lambda_W$ in Equation~\eqref{eq:upLamb}
\begin{equation}
\Lambda_{W.,t+1}\sum_{k\in\mathcal{K}_W}S^{(k)}_t = \sum_{k\in\mathcal{K}_W}\mathbf{X}^{(k)\T}_WM_t^{(k)}
\end{equation}
reduces to 
\begin{equation}\label{eq:foc1inf}
\bar\Lambda_{W.}\sum_{k\in\mathcal{K}_W}
\bar S^{(k)}
 = \sum_{k\in\mathcal{K}_W}n_k \hat\Sigma^{(k)}_{W.}\bar\Sigma^{(k)-1}\bar\Lambda^{(k)}
  \end{equation}
which, by Equation~\eqref{eq:SBAR}, can be rewritten as
\begin{equation}\label{eq:foc1inf3}
\sum_{k\in\mathcal{K}_W}  n_k\left\{ \bar\Lambda_{W.} - \bar\Lambda_{W.}\bar\Lambda^{(k)\T}\bar U^{(k)}\bar\Lambda^{(k)}-\hat\Sigma^{(k)}_{W.}\bar\Sigma^{(k)-1}\bar\Lambda^{(k)}\right\}=0
\end{equation}
The left-hand side of Equation~\eqref{eq:foc1inf3} equals
\begin{eqnarray*}
   && \sum_{k\in\mathcal{K}_W}  n_k\left\{ \bar\Lambda_{W.} - (\bar\Sigma^{(k)}_{W.}-\bar\Psi^{(k)}_{W.})\bar U^{(k)}\bar\Lambda^{(k)}-\hat\Sigma^{(k)}_{W.}\bar\Sigma^{(k)-1}\bar\Lambda^{(k)}\right\}~~~\\
   &=& \sum_{k\in\mathcal{K}_W}  n_k\left\{ \bar\Lambda_{W.} - \bar\Sigma^{(k)}_{W.}\bar\Sigma^{(k)-1}\bar\Lambda^{(k)}   +\bar\Sigma^{(k)}_{W.}\bar\Sigma^{(k)-1}\hat\Sigma^{(k)}\bar\Sigma^{(k)-1}\bar\Lambda^{(k)}\right.\\
   && \left.
   ~~~~~~~~~~~~+\bar\Psi^{(k)}_{W.}\bar U^{(k)}\bar\Lambda^{(k)}-\hat\Sigma^{(k)}_{W.}\bar\Sigma^{(k)-1}\bar\Lambda^{(k)}
   \right\}~~~\\
   &=& \sum_{k\in\mathcal{K}_W}  n_k\left\{ \bar\Lambda_{W.} - I_{W.}\bar\Lambda^{(k)}   +I_{W.}\hat\Sigma^{(k)}\bar\Sigma^{(k)-1}\bar\Lambda^{(k)} 
   +\bar\Psi^{(k)}_{W.}\bar U^{(k)}\bar\Lambda^{(k)}-\hat\Sigma^{(k)}_{W.}\bar\Sigma^{(k)-1}\bar\Lambda^{(k)}
   \right\}~~~\\
   &=& \sum_{k\in\mathcal{K}_W}  n_k\left\{  \bar\Psi^{(k)}_{W.}\bar U^{(k)}\bar\Lambda^{(k)}
   \right\}~~~\\
   &=& \bar\Psi_{WW}\sum_{k\in\mathcal{K}_W}  n_k[\bar U^{(k)}]_{W.}\bar\Lambda^{(k)}
\end{eqnarray*}
where in the first step we applied Lemma~\ref{lemma:matprod}(i), for which
$ \bar\Lambda_{W.}\bar\Lambda^{(k)\T} = [\bar\Lambda^{(k)}\bar\Lambda^{(k)\T}]_{W.}=[ \bar\Sigma^{(k)}-\bar\Psi^{(k)}]_{W.}=\bar\Sigma^{(k)}_{W.}-\bar\Psi^{(k)}_{W.}$; in the third step, we applied Lemma~\ref{lemma:matprod}(ii), for which $\bar\Sigma_{W.}^{(k)}\bar\Sigma^{(k)-1}=I_{W.}$; in the fourth step, we applied Lemma~\ref{lemma:matprod}(i), for which $I_{W.}\bar\Lambda^{(k)}=\bar\Lambda_{W.}$ and $I_{W.}\hat\Sigma^{(k)}=\hat\Sigma^{(k)}_{W.}$; and in the final step, we applied Lemma~\ref{lemma:matprod}(iii), for which $\bar\Psi^{(k)}_{W.}\bar\Sigma^{(k)-1}=\bar\Psi^{(k)}_{WW}\bar\Sigma^{(k)-1}_{W.}=\bar\Psi_{WW}\bar\Sigma^{(k)-1}_{W.}$. Therefore, since $\bar\Psi\succ 0$, we have that for any $W=W_1,\ldots, W_J$, $(\bar\Lambda,\bar\Psi)$ satisfy
\begin{equation}\label{eq:foc1inf2}
\sum_{k\in\mathcal{K}_W}  n_k[\bar U^{(k)}]_{W.}\bar\Lambda^{(k)}=0
\end{equation}
i.e., $(\bar\Lambda,\bar\Psi)$ satisfies the MLE first order condition $\frac{\partial\ell}{\partial\Lambda}=0$.

We now show that if $(\bar\Lambda,\bar\Psi)$ is a converging solution of the EM algorithm, then it also satisfies the first order condition $\frac{\partial\ell}{\partial\Psi}=0$. As $t\to\infty$, the updating Equation~\eqref{eq:upPsiApp} reduces to
\[
\bar\Psi_{WW} = {\rm Diag}\left\{\sum\limits_{k\in\mathcal{K}_{W}}\left(\mathbf{X}_{W}^{(k)\T}\mathbf{X}^{(k)}_{W} -\bar\Lambda_{W.}\bar S^{(k)}\bar\Lambda_{W.}^\T \right)\right\}\left(\sum\limits_{k\in\mathcal{K}_{W}}n_k\right)^{-1},
\]
and by  Equations~\eqref{eq:sigmahatk} and \eqref{eq:foc1inf}, we obtain
\[
\bar\Psi_{WW}\sum\limits_{k\in\mathcal{K}_{W}}n_k = {\rm Diag}\left\{\sum\limits_{k\in\mathcal{K}_{W}}n_k\left(\hat\Sigma^{(k)}_{WW} -\hat\Sigma^{(k)}_{W.}\bar\Sigma^{(k)-1}\bar\Lambda^{(k)}\bar\Lambda_{W.}^\T \right)\right\},
\]
which is equivalent to
\begin{equation}\label{eq:consx}
{\rm Diag}\left\{\sum\limits_{k\in\mathcal{K}_{W}}n_k\left[\bar\Psi^{(k)}-\hat\Sigma^{(k)}  +\hat\Sigma^{(k)}\bar\Sigma^{(k)-1}\bar\Lambda^{(k)}\bar\Lambda^{(k)\T} \right]_{WW}\right\}=0,
\end{equation}
Note that
\begin{eqnarray*}
\bar\Psi^{(k)}-\hat\Sigma^{(k)} +\hat\Sigma^{(k)}\bar\Sigma^{(k)-1}\bar\Lambda^{(k)}\bar\Lambda^{(k)\T} &=& \bar\Psi^{(k)}-\hat\Sigma^{(k)} +\hat\Sigma^{(k)}\bar\Sigma^{(k)-1}(\bar\Sigma^{(k)}-\bar\Psi^{(k)}) \\
&=& \bar\Psi^{(k)}-\hat\Sigma^{(k)}\bar\Sigma^{(k)-1}\bar\Psi^{(k)} \\
&=& \bar\Sigma^{(k)}\bar\Sigma^{(k)-1}\bar\Psi^{(k)}-\bar\Sigma^{(k)}\bar\Sigma^{(k)-1}\hat\Sigma^{(k)}\bar\Sigma^{(k)-1}\bar\Psi^{(k)} \\
&=& \bar\Sigma^{(k)}\bar U^{(k)}\bar\Psi^{(k)} \\
&=& (\bar\Lambda^{(k)}\bar\Lambda^{(k)\T}+\bar\Psi^{(k)})\bar U^{(k)}\bar\Psi^{(k)}
\end{eqnarray*}
and
\begin{eqnarray*}
[(\bar\Lambda^{(k)}\bar\Lambda^{(k)\T}+\bar\Psi^{(k)})\bar U^{(k)}\bar\Psi^{(k)}]_{WW} &=& \left[\bar\Lambda^{(k)}\bar\Lambda^{(k)\T}\bar U^{(k)}+\bar\Psi^{(k)}\bar U^{(k)}\right]_{WW}\bar\Psi_{WW}\\
&=& \left(\bar\Lambda_{W.}\bar\Lambda^{(k)\T}\bar U^{(k)}_{.W}+\bar\Psi_{WW}\bar U^{(k)}_{WW}\right)\bar\Psi_{WW},
\end{eqnarray*}
where $\bar U^{(k)}$ is defined in Equation~\eqref{eq:ubar}. 
Thus, since $\bar\Psi_{WW}$ is a positive diagonal matrix,  Equation~\eqref{eq:consx} is equivalent to
\[
{\rm Diag}\left\{\bar\Lambda_{W.}\sum\limits_{k\in\mathcal{K}_{W}} n_k\bar\Lambda^{(k)\T}\bar U^{(k)}_{.W}
+\bar\Psi_{WW}\sum\limits_{k\in\mathcal{K}_{W}} n_k\bar U^{(k)}_{WW}\right\}=0
\]
where, by Equation~\eqref{eq:foc1inf2}, 
\[
\sum_{k\in\mathcal{K}_W} n_k \bar\Lambda^{(k)\T}\bar U^{(k)}_{.W} = 0
\]
Therefore, Equation~\eqref{eq:consx} is equivalent to
\[ \sum\limits_{k\in\mathcal{K}_{W}} n_k {\rm Diag}\left\{\bar U^{(k)}_{WW}\right\}=0
\]
for any $W=W_1,\ldots, W_J$,  i.e., the converging solution $(\bar\Lambda,\bar\Psi)$ satisfies the MLE first order condition $\frac{\partial\ell}{\partial\Psi}=0$.
\end{proof}

\subsection{Statistical properties of the LINFA MLE}\label{app:proofstheory}

\begin{proof}[Proof of Theorem~\ref{theo:consistency}]
Let $\theta_0\in\tilde{\mathcal{P}}_b:=\{ ({\rm vec}(\Lambda),{\rm diag}(\Psi))^\T: (\Lambda,\Psi)\in\mathcal{P}_b)\}$ be the true value of the parameter $\theta$,  where $\mathcal{P}_b$ is given in Equation~\eqref{eq:theoset}, and let $\hat\theta\in\tilde{\mathcal{P}}_b$ be the constrained MLE of $\theta_0$. 
Define the random function
\begin{eqnarray}
    M_n(\theta) &:=& \tfrac{1}{n}\left(\ell_n(\theta_0)-\ell_n(\theta)\right)\\
    &=& \tfrac{1}{n}\sum_{k=1}^K \left(\ell^{(k)}_{n_k}(\theta_0^{(k)})-\ell^{(k)}_{n_k}(\theta^{(k)})\right)\\
        &=& \sum_{k=1}^K \tfrac{n_k}{n}\tfrac{1}{n_k} \left(\ell^{(k)}_{n_k}(\theta_0^{(k)})-\ell^{(k)}_{n_k}(\theta^{(k)})\right)\\
        &=& \sum_{k=1}^K \tfrac{n_k}{n}M_{n_k}^{(k)}(\theta^{(k)})
\end{eqnarray}
where $\ell_n(\cdot)$ is the LINFA log-likelihood function,
\begin{equation}
M_{n_k}^{(k)}(\theta^{(k)}):=  \tfrac{1}{n_k} \sum_{r\in\mathcal{N}_k}\log\tfrac{f_k(X^{(k)}_r;\theta_0^{(k)})}{f_k(X^{(k)}_r;\theta^{(k)})},
\end{equation}
$f_k$ is the marginal p.d.f.~of $X_r^{(k)}$, and $\theta^{(k)}=({\rm vec}(\Lambda^{(k)})^\T,{\rm diag}(\Psi^{(k)})^\T)^\T$. Note that
\begin{eqnarray}
M_n(\theta)&\ge& M_n(\hat\theta),~\forall\theta\in\tilde{\mathcal{P}}_b,\\
 M_n(\hat\theta)&\le & M_n(\theta_0) = 0
\end{eqnarray}
Moreover, define the Kullback-Leibler divergence
\begin{eqnarray}
\mathcal{D}_k(\theta_0^{(k)}\Vert\theta^{(k)}) &:=& \E_{\theta_0}\left[\log\tfrac{f_k(X^{(k)}_r;\theta_0^{(k)})}{f_k(X^{(k)}_r;\theta^{(k)})}\right]\\
&=&\tfrac{1}{2}\E_{\theta_0}\left[\log\tfrac{\det\Sigma^{(k)}}{\det\Sigma^{(k)}_0}+{\rm tr}\left((\Sigma^{(k)-1}-\Sigma_0^{(k)-1})X_r^{(k)}X_r^{(k)\T}\right)\right]\\
&=& \tfrac{1}{2}\left( \log\tfrac{\det\Sigma^{(k)}}{\det\Sigma^{(k)}_0}-|V_k|+{\rm tr}\left(\Sigma^{(k)-1}\Sigma^{(k)}_0\right)\right)
\end{eqnarray}
Note that, if $\theta\in\tilde{\mathcal{P}}_b$, we have $\lambda_{\min}(\Sigma),\lambda_{\max}(\Sigma)^{-1}\ge b>0$ and thereby, by Cauchy's Interlace Theorem, $\lambda_{\rm min}(\Sigma^{(k)}),\lambda_{\rm max}(\Sigma^{(k)})^{-1}\ge b>0$, for all $k=1,\ldots,K$. Thus, $\forall\theta\in\tilde{\mathcal{P}}_b$,
\begin{eqnarray}\label{eq:unifbound1}
&&\left|\log\tfrac{f_k(X^{(k)}_r;\theta_0^{(k)})}{f_k(X^{(k)}_r;\theta^{(k)})}\right|  \\
&=& \tfrac{1}{2}\left|\log\tfrac{\det\Sigma^{(k)}}{\det\Sigma^{(k)}_0}+{\rm tr}\left((\Sigma^{(k)-1}-\Sigma_0^{(k)-1})X_r^{(k)}X_r^{(k)\T}\right)\right|\\
&\le &\tfrac{1}{2}\left|\log\tfrac{\det\Sigma^{(k)}}{\det\Sigma^{(k)}_0}\right| + \tfrac{1}{2}\left|{\rm tr}\left(\Sigma^{(k)-1} X_r^{(k)}X_r^{(k)\T}\right)\right|+ \tfrac{1}{2}\left|{\rm tr}\left(\Sigma_0^{(k)-1} X_r^{(k)}X_r^{(k)\T}\right)\right|\\
&\le & \tfrac{1}{2}\max\left(\log \tfrac{b^{-|V_k|}}{b^{|V_k|}},~-\log \tfrac{b^{|V_k|}}{b^{-|V_k|}}\right) + \tfrac{1}{2}\left(b^{-1}+  b^{-1}\right){\rm tr}(X_r^{(k)}X_r^{(k)\T})~~~~~~~~~~~\\
&=& -|V_k|\log b+b^{-1}X_r^{(k)\T}X_r^{(k)}\\
&=: & h(X_r^{(k)}),
\end{eqnarray}
where, in the last inequality, we used the inequality ${\rm tr}(AB)\le \lambda_{\max}(A)\Vert B\Vert^2_F$ for positive semi-definite matrices $A,B$. 
Consequently,
\begin{eqnarray}\label{eq:unifbound2}
\E\left[\left|\log\tfrac{f_k(X^{(k)}_r;\theta_0^{(k)})}{f_k(X^{(k)}_r;\theta^{(k)})}\right|\right]&\le&\E[h(X_r^{(k)})]\\
&=& -|V_k|\log b+b^{-1}{\rm tr}(\Sigma_0^{(k)})\\
&\le& -|V_k|\log b+|V_k|b^{-2}\\
&=& |V_k|(b^{-2}-\log b)\\
&<& \infty
\end{eqnarray}
Thus, by the Weak Law of Large Numbers, $\forall\theta\in\tilde{\mathcal{P}}_b$, as $n\to\infty$,
\begin{eqnarray}
  M_{n_k}^{(k)}(\theta^{(k)}) &\pto &  \mathcal{D}_k(\theta_0^{(k)}\Vert\theta^{(k)})
\end{eqnarray}
and since $\frac{n_k}{n}=\frac{\lfloor n\cdot p_k\rceil}{n}\to p_k$,
\begin{eqnarray}\label{eq:Mpointconv}
    M_n(\theta)\pto M(\theta) &:=& \sum_{k=1}^K p_k \mathcal{D}_k(\theta_0^{(k)}\Vert\theta^{(k)})
\end{eqnarray}
Moreover, $M_n(\theta)$ also converges to $M(\theta)$ \textit{uniformly} in probability. Indeed,
 \begin{eqnarray}
\sup\limits_{\theta\in\tilde{\mathcal{P}}_b} |M_n(\theta)-M(\theta)| &=& \sup\limits_{\theta\in\tilde{\mathcal{P}}_b} \left|\sum_{k=1}^K \tfrac{n_k}{n}M_{n_k}^{(k)}(\theta^{(k)})- \sum_{k=1}^K p_k \mathcal{D}_k(\theta_0^{(k)}\Vert\theta^{(k)})\right|\\
&=& \sup\limits_{\theta\in\tilde{\mathcal{P}}_b} \left|\sum_{k=1}^K \tfrac{n_k}{n}\left( M_{n_k}^{(k)}(\theta^{(k)})-  \tfrac{np_k}{n_k}\mathcal{D}_k(\theta_0^{(k)}\Vert\theta^{(k)})\right)\right|~~~~~~~\\
&\le & \sum_{k=1}^K \tfrac{n_k}{n}\sup\limits_{\theta\in\tilde{\mathcal{C}}} \left| M_{n_k}^{(k)}(\theta^{(k)})-  \tfrac{np_k}{n_k}\mathcal{D}_k(\theta_0^{(k)}\Vert\theta^{(k)}) \right|\\
&\le & \sum_{k=1}^K \tfrac{n_k}{n}\sup\limits_{\theta\in\tilde{\mathcal{P}}_b} \left| M_{n_k}^{(k)}(\theta^{(k)})-\mathcal{D}_k(\theta_0^{(k)}\Vert\theta^{(k)})\right|\\
&&+\sum_{k=1}^K \tfrac{n_k}{n}\sup\limits_{\theta\in\tilde{\mathcal{P}}_b} \left|\mathcal{D}_k(\theta_0^{(k)}\Vert\theta^{(k)})-  \tfrac{np_k}{n_k}\mathcal{D}_k(\theta_0^{(k)}\Vert\theta^{(k)}) \right|
\end{eqnarray}
By Equation~\eqref{eq:unifbound2}, 
\begin{equation} \sup\limits_{\theta\in\tilde{\mathcal{P}}_b} \left| M_{n_k}^{(k)}(\theta^{(k)})-\mathcal{D}_k(\theta_0^{(k)}\Vert\theta^{(k)})\right|\pto 0
\end{equation}
and since $\frac{n_k}{n}\to p_k$ and $0\le\mathcal{D}_k(\theta_0^{(k)}\Vert\theta^{(k)})\le \E[h(X_r^{(k)})]\le |V_k|(b^{-2}-\log b)<\infty,\forall\theta\in\tilde{\mathcal{P}}_b$, we obtain 
\begin{eqnarray}
\sup\limits_{\theta\in\tilde{\mathcal{P}}_b} \left|\mathcal{D}_k(\theta_0^{(k)}\Vert\theta^{(k)})-  \tfrac{np_k}{n_k}\mathcal{D}_k(\theta_0^{(k)}\Vert\theta^{(k)}) \right|&=&\left|1-\tfrac{np_k}{n_k}\right|\cdot\sup\limits_{\theta\in\tilde{\mathcal{P}}_b}  \mathcal{D}_k(\theta_0^{(k)}\Vert\theta^{(k)}) \\  &\le & \left|1-\tfrac{np_k}{n_k}\right|\cdot|V_k|(b^{-2}-\log b) ~\to~0
\end{eqnarray}
Therefore, 
\begin{equation}\label{eq:Munifconv}
\sup\limits_{\theta\in\tilde{\mathcal{P}}_b} |M_n(\theta)-M(\theta)|\pto 0
\end{equation}
If the sets $V_1,\ldots,V_K$ are $q$-linked, the matrices $\Sigma_0^{(1)},\ldots,\Sigma_0^{(K)}$ are uniquely identified by $\theta_0$, so 
\begin{equation}\label{eq:strongident1}
M(\theta)\ge M(\theta_0)=0, \forall\theta\in\tilde{\mathcal{P}}_b
\end{equation}
with equality if and only if $\theta=\theta_0$. Therefore, by Equations~\eqref{eq:Mpointconv}, \eqref{eq:Munifconv}, and \eqref{eq:strongident1}, 
\begin{eqnarray}
    0 ~\le~ M(\hat\theta)-M(\theta_0) &=& M(\hat\theta)-M_n(\theta_0)+M_n(\theta_0)-M(\theta_0)\\
    &\le & M(\hat\theta)-M_n(\hat\theta)+M_n(\theta_0)-M(\theta_0)\\
    &\le & \sup_{\theta\in\tilde{\mathcal{P}}_b} |M_n(\theta)-M(\theta)|+M_n(\theta_0)-M(\theta_0)\\
    &\pto& 0
\end{eqnarray}
so, for any $\delta>0$, as $n\to\infty$,
\begin{eqnarray}
    P\left(M(\hat\theta)>M(\theta_0)+\delta\right)\to 0
\end{eqnarray}
By Equation~\eqref{eq:strongident1}
we have that, for all $\epsilon>0$,
\[
\inf_{\theta\in\tilde{\mathcal{P}_b}:~{\rm dist}(\theta,\theta_0)\ge\epsilon}~M(\theta)>0
\] 
where ${\rm dist}(\theta,\theta_0)$ is a distance metric (e.g., ${\rm dist}(\theta,\theta_0)=\Vert \hat\theta-\theta_0\Vert$). Since $M(\theta)=0$ if and only if ${\rm dist}(\theta,\theta_0)=0$, we have that, for any $\epsilon>0$, there exists $\delta>0$ such that ${\rm dist}(\theta,\theta_0)\ge\epsilon$ implies $M(\theta)> M(\theta_0)+\delta$. Therefore, for all $\epsilon>0$,
\begin{eqnarray}
    0 &\le &P({\rm dist}(\hat\theta,\theta_0)>\epsilon)\\
    &\le& P({\rm dist}(\hat\theta,\theta_0)\ge\epsilon)\\
    &\le& P\left(M(\hat\theta)>M(\theta_0)+\delta\right)\to 0
\end{eqnarray}
that is, ${\rm dist}(\hat\theta,\theta_0)\pto 0$, which implies 
\begin{eqnarray}
\Vert \hat\Lambda -\Lambda_0\Vert &\pto& 0\\
    \Vert\hat\Psi-\Psi_0\Vert &\pto&0
\end{eqnarray}
\end{proof}

\begin{proof}[Proof of Theorem~\ref{theo:asynorm}]
Define the Lagrangian
\begin{eqnarray}
 \mathcal{L}(\theta,\lambda)&=& 
\ell(\theta)-\lambda^\T g(\theta)
\end{eqnarray}
where
\begin{eqnarray}
 \ell(\theta)&=& \sum_{k=1}^K\ell_k(\theta^{(k)})
\end{eqnarray}
is the LINFA log-likelihood function with $k$-th component $\ell_k$ depending on the parameter subset $\theta^{(k)}=({\rm vec}(\Lambda^{(k)})^\T,{\rm diag}(\Psi^{(k)})^\T)^\T$, 
$\lambda = (\lambda_{ij})_{1\le i<j\le q}$ is a vector of multipliers, and
\begin{eqnarray}
g(\theta)&=& \left(g_{ij}(\theta)\right)_{1\le i<j\le q},
\end{eqnarray}
is a vector with entry
\begin{equation}
g_{ij}(\theta) = [\Lambda^\T\Psi^{-1}\Lambda]_{ij} =  \sum_{h=1}^d\Lambda_{hi}\Lambda_{hj}/\Psi_{hh}
\end{equation}
Let $\hat\theta$ be the MLE of $\theta$ after rotation (Equation~\eqref{eq:rotMLE}) and sign flip correction (Equation~\eqref{eq:signflip}). Then, by Theorem~\eqref{theo:consistency}, 
\begin{equation}
\hat\theta - \theta_0\pto 0
\end{equation}
Moreover, $\hat\theta$ satisfies the constraint $[\Lambda^\T\Psi^{-1}\Lambda]_{ij}=0$ for $1\le i<j\le q$, so there exists a vector $\hat\lambda$ such that 
\begin{eqnarray}
\nabla\mathcal{L}(\hat\theta,\hat\lambda) &=&0
\end{eqnarray}
where
\begin{eqnarray}
\nabla\mathcal{L}(\theta,\lambda) &=&\left[\begin{array}{c}
\sum\limits_{k=1}^K \ell_k'(\theta^{(k)})-  g'(\theta)^\T\lambda\\
-g(\theta)
\end{array}\right]
\end{eqnarray}
is the gradient of $\mathcal{L}$ with respect to $(\theta,\lambda)$,  $\ell_k'$ is the gradient of $\ell_k$ with respect to $\theta$, and $g'$ is the Jacobian matrix of $g$ with respect to $\theta$. 
By taking the first order Taylor expansion of $\nabla\mathcal{L}(\hat\theta,\hat\lambda)$ about the true parameter value $\theta=\theta_0$, we obtain
\begin{eqnarray}
\sum\limits_{k=1}^K \ell_k'(\theta_0^{(k)})- g'(\theta_0)^\T\hat\lambda +\left(   \sum\limits_{k=1}^K\ell_k''(\theta_0^{(k)})- \sum\limits_{1\le i<j\le q}\hat\lambda_{ij} g_{ij}''(\theta_0) \right) (\hat\theta-\theta_0) &\approx& 0\\
g(\theta_0)+g'(\theta_0) (\hat\theta-\theta_0)
    &\approx& 0~~~~~
\end{eqnarray}
where $\ell_k''$ and $g_{ij}''$ are Hessian matrices. By multiplying the first equation by $1/\sqrt{n}$ and the second equation by $\sqrt{n}$, and using the fact $g(\theta_0)=0$, we obtain the system
{\footnotesize \begin{eqnarray*}
    \sum\limits_{k=1}^K \sqrt{\tfrac{n_k}{n}}\tfrac{1}{\sqrt{n_k}}\ell_k'(\theta_0^{(k)})-  \frac{1}{\sqrt{n}}g'(\theta_0)^\T\hat\lambda
    -
    \left(
\sum\limits_{k=1}^K\tfrac{n_k}{n}\left(-\tfrac{1}{n_k} \ell_k''(\theta_0^{(k)})\right)+ \frac{1}{n}\sum\limits_{1\le i<j\le q}\hat\lambda_{ij} g_{ij}''(\theta_0) \right) \sqrt{n}(\hat\theta-\theta_0) &\approx& 0\\
    g'(\theta_0) \sqrt{n}(\hat\theta-\theta_0)
    &\approx& 0~~~~~
\end{eqnarray*}}
or, equivalently, 
{\footnotesize 
\begin{equation*}
 \left[ \begin{array}{cc}
\sum\limits_{k=1}^K\tfrac{n_k}{n}\left(-\tfrac{1}{n_k} \ell_k''(\theta_0^{(k)})\right)+ \frac{1}{n}\sum\limits_{1\le i<j\le q}\hat\lambda_{ij} g_{ij}''(\theta_0), & g'(\theta_0)^\T\\
g'(\theta_0), & 0
    \end{array}
    \right] \times \left[ \begin{array}{c}
\sqrt{n}(\hat\theta-\theta_0)\\
\hat\lambda/\sqrt{n}
    \end{array}
    \right]  \approx \left[ \begin{array}{c}
\sum\limits_{k=1}^K \sqrt{\tfrac{n_k}{n}}\tfrac{1}{\sqrt{n_k}}\ell_k'(\theta_0^{(k)})\\
0
    \end{array}
    \right]
\end{equation*} }
Now, note that, as $n\to\infty$, 
\begin{eqnarray}
\frac{n_k}{n} &\to& p_k,\\
\sqrt{\frac{n_k}{n}} &\to& \sqrt{p_k},\\
\frac{1}{\sqrt{n_k}}\ell_k'(\theta_0^{(k)}) &\dto& N\left(0, \tilde{\mathcal{I}}_k(\theta_0) \right),\\
-\frac{1}{n_k} \ell_k''(\theta_0^{(k)}) &\pto& \tilde{\mathcal{I}}_k(\theta_0),
\end{eqnarray}
where $\tilde{\mathcal{I}}_k(\theta_0)$ is the Fisher Information matrix about $\theta$ contained in the random vector $X_{V_k}$. Moreover,
\begin{eqnarray}
\frac{1}{n}\sum\limits_{1\le i<j\le q}\hat\lambda_{ij} g_{ij}''(\theta_0) &\pto& 0
\end{eqnarray}
since $\hat\lambda\pto \lambda_0$ for some finite constant vector $\lambda_0$ in order to have $\nabla\mathcal{L}(\hat\theta,\hat\lambda)=0$ as $n\to\infty$.  

Thus,
\begin{eqnarray}
\left[ \begin{array}{cc}
\tilde{\mathcal{I}}(\theta_0), & g'(\theta_0)^\T\\
g'(\theta_0), & 0
    \end{array}
    \right]  \times \left[ \begin{array}{c}
\sqrt{n}(\hat\theta-\theta_0)\\
\hat\lambda/\sqrt{n}
    \end{array}
    \right]  & \dto &   N\left(0, ~
    \left[
\begin{array}{cc}
\tilde{\mathcal{I}}(\theta_0), & 0\\
0, & 0
\end{array}
    \right]
    \right),
\end{eqnarray}
where 
\begin{equation}
    \tilde{\mathcal{I}}(\theta_0) := \sum\limits_{k=1}^K p_k \tilde{\mathcal{I}}_k(\theta_0)
\end{equation}
is the \textit{weighted Fisher Information Matrix}. 
Therefore,
\begin{equation}
    \sqrt{n}(\hat\theta-\theta_0) \dto N(0,A\tilde{\mathcal{I}}(\theta_0)A^\T)
\end{equation}
where 
\begin{eqnarray}
    A &=& \left(\left[ \begin{array}{cc}
\tilde{\mathcal{I}}(\theta_0), & g'(\theta_0)^\T\\
g'(\theta_0), & 0
    \end{array}
    \right]^{-1}\right)_{1:d(q+1),1:d(q+1)}
\end{eqnarray}
\end{proof}

\begin{proof}[Proof of Corollary~\ref{coro:sigma}]
Define the vector valued function $h:\mathbb{R}^{d(q+1)}\to \mathbb{R}^{d^2}$ as  $h(\theta) = {\rm vec}(\Sigma) =  {\rm vec}(\Lambda\Lambda^\T+\Psi)$. By the Multivariate Delta Method, 
\begin{eqnarray}
\sqrt{n}~{\rm vec}(\hat\Sigma-\Sigma) = \sqrt{n}~(h(\hat\theta)-h(\theta))\dto N(0,J\Omega J)
\end{eqnarray}
where  $J = \frac{\partial}{\partial\theta}h(\theta)$ is the $d^2\times (d(q+1))$ Jacobian matrix of $h$. Precisely, we have
\begin{eqnarray}
    J = [\Lambda\otimes I_d+I_d\otimes\Lambda\times C^{(d,q)},S]
\end{eqnarray}
where 
$C^{(d,q)}$ is a commutation matrix, $S$ is a $d^2\times d^2$ matrix with the $j$-th column equal to ${\rm vec}(E^{(j)})$,  $E^{(j)}$ is a matrix with entry $ E^{(j)}_{lk}=I(l=k=j)$, and $I()$ is the indicator function.
\end{proof}

\begin{proof}[Proof of Theorem~\ref{theo:confreg}] Let $(\Lambda_0,\Psi_0)$ be the true parameter values. To show that $\mathcal{C}_n$ is an approximate $1-\alpha$ confidence region for $(\Lambda_0,\Psi_0)$, we need to show that, as $n\to\infty$,
\begin{equation}
\lambda_n ~:=~ 2(\ell_n(\hat\Lambda,\hat\Psi)-\ell_n(\Lambda_0,\Psi_0))~\dto~ \chi^2_{\kappa}
\end{equation}
so that $P(\lambda_n\le\chi^2_{\kappa,\alpha})\approx 1-\alpha$ for large $n$, 

Let $\vartheta$ be the vector of effectively free parameters after applying the canonical constraints (Equation~\eqref{eq:canon}) on $\Lambda$ and $\Psi$. Specifically, $\vartheta$ contains all the $d$ diagonal entries of $\Psi$, and an (arbitrary) subset of $dq-q(q-1)/2$ entries of $\Lambda$, where $q(q-1)/2$ is the number of constraints in Equation~\eqref{eq:canon}. Thus, the number of dimensions of $\vartheta$ is 
\begin{equation}\label{eq:kappapp}
\kappa=d(q+1)-q(q-1)/2
\end{equation}
Let $\vartheta_0$ be the true value of $\vartheta$, and let $\hat\vartheta$ be the MLE. 
Thus, we can rewrite
\begin{eqnarray}
\lambda_n &:=& 2(\ell_n(\hat\vartheta)-\ell_n(\vartheta_0)) 
\end{eqnarray}
where $\ell_n(\vartheta)$ is the LINFA loglikelihood function reparametrized in terms of $\vartheta$.

Since $\hat\vartheta\pto\vartheta_0$ by Theorem~\ref{theo:consistency}, for large $n$, the following approximation based on the second order
Taylor expansion of $\ell(\vartheta_0)$ about $\vartheta=\hat\vartheta$ holds:
\begin{eqnarray}
\ell(\vartheta_0) &=&  \ell(\hat\vartheta)+(\vartheta_0-\hat\vartheta)^\T \nabla\ell(\hat\vartheta)+\frac{1}{2}(\vartheta_0-\hat\vartheta)^\T H(\hat\vartheta)  (\vartheta_0-\hat\vartheta)+o_p(1) ~~~~~~
\end{eqnarray}
where $\nabla\ell(\hat\vartheta)=0$ and 
\begin{eqnarray}
 H(\hat\vartheta) ~=~  \sum_{k=1}^K\left. n_k\frac{\partial^2 \ell_k(\vartheta^{(k)})}{\partial\vartheta\partial\vartheta^\T}\right\vert_{\vartheta=\hat\vartheta}
\end{eqnarray}
is the Hessian matrix of $\ell(\vartheta)$ evaluated at $\hat\vartheta$. By rearranging terms, we obtain
\begin{eqnarray}
\lambda_n &=& \sqrt{n}(\hat\vartheta-\vartheta_0)^\T \left(-n^{-1}H(\hat\vartheta)\right)  \sqrt{n}(\hat\vartheta-\vartheta_0)+o_p(1) 
\end{eqnarray}
where $n^{-1}H(\hat\vartheta) $ 
is the observed Fisher Information Matrix about $\vartheta$ and, by the Weak Law of Large Numbers,
\begin{eqnarray}
    -n^{-1}H(\hat\vartheta)\pto B,
\end{eqnarray}
where $B$ is the $\kappa\times \kappa$ submatrix of the Fisher Information Matrix $\mathcal{I}(\theta)$ (Equation~\eqref{eq:fishinfo}) relative to the components in $\vartheta$. The submatrix $B$ has rank $\kappa$ and is invertible because the components of $\vartheta$ are free parameters. Since $\sqrt{n}(\hat\vartheta-\vartheta_0)\dto N(0,B^{-1})$ by Theorem~\ref{theo:asynorm}, by the Continuous Mapping Theorem and Slutsky's Theorem we obtain
\begin{eqnarray}
Z_n:=\sqrt{n}(\hat\vartheta-\vartheta_0)^\T \left(-n^{-1}H(\hat\vartheta)\right)  \sqrt{n}(\hat\vartheta-\vartheta_0) &\dto& \chi^2_{\kappa}
\end{eqnarray}
Finally, by Slutsky's Theorem, $\lambda_n=Z_n+o_p(1)\dto \chi^2_{\kappa}$.
\end{proof}

\begin{proof}[Proof of Corollary~\ref{coro:ht}]
Note that, $(\Lambda_0,\Psi_0)\notin\mathcal{C}$ is equivalent to 
\[
\lambda_n:=2(\ell_n(\hat\Lambda,\hat\Psi)-\ell_n(\Lambda_0,\Psi_0))
> \chi^2_{\kappa,\alpha}
\]
which, under $H_0$, happens with probability $\approx \alpha$, i.e., the probability of type I error is approximately $\alpha$.
\end{proof}

\subsection{Model selection}\label{app:modelsel}

\begin{theorem}[\sc Bayesian Information Criterion for LINFA]\label{theo:BIC}
The Bayesian Information Criterion risk estimator for LINFA based on the likelihood function in Equation~\eqref{eq:linfalikelihood} and canonical rotation (Equation~\eqref{eq:canon}) is
\begin{equation}
    \textsc{Risk}_{\rm BIC}(q) ~=~ -2\log L(\hat\Lambda,\hat\Psi;\mathbf{X}^{(1)},\ldots,\mathbf{X}^{(K)})+\left(d(q+1)-q(q-1)/2\right)\log n
\end{equation}
\begin{proof}

Let $\vartheta$ be the vector of effectively free parameters after applying the canonical constraints (Equation~\eqref{eq:canon}) on $\Lambda$ and $\Psi$. Note that $\vartheta$ contains all the $d$ diagonal entries of $\Psi$, and an (arbitrary) subset of $dq-q(q-1)/2$ entries of $\Lambda$, where $q(q-1)/2$ is the number of constraints in Equation~\eqref{eq:canon}. Thus, the number of dimensions of $\vartheta$ is 
\begin{equation}
\kappa=d(q+1)-q(q-1)/2
\end{equation}
Given some prior distribution $\pi(\vartheta)$, the marginal likelihood of the data $\mathbf{X}^{(1)},\ldots,\mathbf{X}^{(K)}$ is
\begin{eqnarray}
g(\mathbf{X}^{(1)},\ldots,\mathbf{X}^{(K)}) &=& \int_\Theta L(\vartheta; \mathbf{X}^{(1)},\ldots, \mathbf{X}^{(K)}) \pi(\vartheta)d\vartheta\\
&=& \int_\Theta \exp\left\{\ell(\vartheta)\right\}\pi(\vartheta) d\vartheta\label{eq:marglik}
\end{eqnarray}
where $\ell(\vartheta):= \sum_{k=1}^K\ell_k(\vartheta^{(k)})$,  $\ell_k(\vartheta^{(k)})=\log L_k(\vartheta^{(k)};\mathbf{X}^{(k)})$, and $\vartheta^{(k)}$ is the subset of parameters that affect $\ell_k$. We now approximate the expression in Equation~\eqref{eq:marglik} via Laplace's Method. Since $\hat\vartheta\pto\vartheta$ by Theorem~\ref{theo:consistency}, the second order Taylor expansion of $\ell(\vartheta)$ about the MLE $\hat\vartheta$ is
\begin{eqnarray}
\ell(\vartheta) &=& \ell(\hat\vartheta)-\frac{1}{2}(\vartheta-\hat\vartheta)^\T\times \left(-H(\hat\vartheta)\right)\times(\vartheta-\hat\vartheta)+o_p(1)
\end{eqnarray}
where we used the fact $\nabla\ell(\hat\vartheta)=0$, and where
\begin{eqnarray}
-\frac{1}{n}H(\hat\vartheta) ~=~ -\frac{1}{n}\sum_{k=1}^K\left. n_k\frac{\partial^2 \ell_k(\vartheta^{(k)})}{\partial\vartheta\partial\vartheta^\T}\right\vert_{\vartheta=\hat\vartheta}
\end{eqnarray}
is the $\kappa\times\kappa$  \textit{observed Fisher Information Matrix} about $\vartheta$. This matrix is positive definite, hence invertible. The first order Taylor expansion of $\pi(\vartheta)$ about the MLE $\hat\vartheta$ is
\begin{eqnarray}
\pi(\vartheta) ~=~ \pi(\hat\vartheta)+(\vartheta-\hat\vartheta)^\T\nabla\pi(\hat\vartheta) +o_p(1)
\end{eqnarray}
Thus, Equation~\eqref{eq:marglik} can be approximated as 
\begin{eqnarray*}
g(\mathbf{X}^{(1)},\ldots,\mathbf{X}^{(K)}) &\approx & \exp\{\ell(\hat\vartheta)\}\pi(\hat\vartheta)\int_\Theta \exp\left(-\frac{1}{2}(\vartheta-\hat\vartheta)^\T\left(-H(\hat\vartheta)\right)(\vartheta-\hat\vartheta)\right)d\vartheta
\\ 
&\approx & \exp\{\ell(\hat\vartheta)\}\pi(\hat\vartheta)\int_{\mathbb{R}^{\kappa}} \exp\left(-\frac{1}{2}(\vartheta-\hat\vartheta)^\T\left(-H(\hat\vartheta)\right)(\vartheta-\hat\vartheta)\right)d\vartheta
\\ 
&=& \exp\{\ell(\hat\vartheta)\}\pi(\hat\vartheta)(2\pi)^{\kappa/2}\left(\det\left(-H(\hat\vartheta)\right)\right)^{-1/2}\\
&=& \exp\{\ell(\hat\vartheta)\}\pi(\hat\vartheta)(2\pi)^{\kappa/2}n^{-\kappa/2}\left(\det\left(-n^{-1}H(\hat\vartheta)\right)\right)^{-1/2}
\end{eqnarray*}
By the Weak Law of Large Numbers,
\begin{eqnarray}
    -n^{-1}H(\hat\vartheta)\pto B,
\end{eqnarray}
where $B$ is the $\kappa\times \kappa$ submatrix of the Fisher Information Matrix $\mathcal{I}(\theta)$ (Equation~\eqref{eq:fishinfo}) relative to the components in $\vartheta$. Thus, $\det\left(-n^{-1}H(\hat\vartheta)\right)=O_p(1)$. 
Therefore, for large $n$, 
\begin{eqnarray*}
g(\mathbf{X}^{(1)},\ldots,\mathbf{X}^{(K)}) &\approx & \exp\left\{\ell(\hat\vartheta)-\tfrac{\kappa}{2}\log n+O_p(1)\right\}\\
&=& \exp\left\{-\frac{\textsc{Risk}_{\rm BIC}(q)}{2}+O_p(1)\right\}
\end{eqnarray*}
where 
\begin{equation}\label{eq:bicproof}
\textsc{Risk}_{\rm BIC}(q) ~=~ -2\ell(\hat\vartheta)+\kappa\log n
\end{equation}
is the Bayesian Information Criterion risk metric.
\end{proof}
\end{theorem}

~

\section{Lemmas}\label{app:lemmas}
This appendix contains important lemmas and their proofs.

\begin{lemma}[\sc Canonical Rotation]\label{lemma:canonrot}
If $(\tilde\Lambda,\tilde\Psi)$ is a maximizer of the likelihood function in Equation~\eqref{eq:linfalikelihood} and $\tilde\Lambda^\T\tilde\Psi^{-1}\tilde\Lambda$ has eigendecomposition $BAB^\T$ with positive diagonal matrix of eigenvalues $A$ and matrix of eigenvectors $B$, then $(\breve\Lambda,\tilde\Psi)$ with $\breve\Lambda=\tilde\Lambda B$ is also a maximizer of the likelihood function in Equation~\eqref{eq:linfalikelihood} and $\breve\Lambda^\T\tilde\Psi^{-1}\breve\Lambda$ is a positive diagonal matrix with diagonals in decreasing order. 
\begin{proof}
Since $B$ is a rotation matrix, we have $BB^\T=I$. Thus,
\[
\breve\Sigma=\breve\Lambda\breve\Lambda^\T+\tilde\Psi=\tilde\Lambda BB^\T\tilde\Lambda^\T+\tilde\Psi=\tilde\Sigma,
\]
which ensures $(\breve\Lambda,\tilde\Psi)$ to be a maximizer of the likelihood function in Equation~\eqref{eq:linfalikelihood}, and
\[
\breve \Lambda^\T\Psi^{-1}\breve \Lambda=B^\T \tilde\Lambda^\T\Psi^{-1}\tilde\Lambda B= B^\T BAB^\T B = A,
\]
where $A$ is a positive diagonal matrix with diagonals $A_{11}>\ldots>A_{qq}$.
\end{proof}
\end{lemma}

\begin{lemma}[\sc Conditional expectations in Factor Model]\label{lemma:condexpEM}
Suppose $X=\Lambda Z+\varepsilon$, where $\Lambda\in\mathbb{R}^{d\times q}$, $d>q$,  $\varepsilon\sim N(0,\Psi)$, and $\Psi$ is a $d\times d$ positive diagonal matrix. Then
\begin{equation}
  \E[Z\mid X]~=~ \Lambda^\T \Sigma^{-1}X
\end{equation}
and
\begin{equation}
 \E[Z^\T Z\mid X]= (I_q-\Lambda^\T\Sigma^{-1}\Lambda)+\Lambda^\T \Sigma^{-1}XX^\T\Sigma^{-1}\Lambda
\end{equation}
where $\Sigma = \Lambda\Lambda^\T+\Psi$.
\begin{proof}
We have $(X^\T,Z^\T)^\T\sim N(0,H)$, where
\[
    H :=~ \cov\left(\left[\begin{array}{c}
         X\\
         Z
    \end{array}\right]\right) 
    ~=~ \left[
    \begin{array}{cc}
       \Sigma  & \Lambda \\
       \Lambda^\T  & I_q
    \end{array}
    \right]
\]
So, by standard properties of the multivariate Gaussian distribution, we obtain
\begin{eqnarray*}
    \E[Z\mid X] &=& 0+H_{ZX}H_{XX}^{-1}X\\
        &=& \Lambda^\T \Sigma^{-1}X
\end{eqnarray*}
and
\begin{eqnarray*}
    \cov(Z\mid X) &=& H_{ZZ}-H_{ZX}H_{XX}^{-1}H_{XZ}\\
                &=& I_q-\Lambda^\T\Sigma^{-1}\Lambda
\end{eqnarray*}
so
\begin{eqnarray*}
     \E[Z^\T Z\mid X] &=&  \cov(Z\mid X)+\E[Z\mid X]\E[Z\mid X]^\T\\
     &=& (I_q-\Lambda^\T\Sigma^{-1}\Lambda)+\Lambda^\T \Sigma^{-1}XX^\T\Sigma^{-1}\Lambda
\end{eqnarray*}
\end{proof}
\end{lemma}

\begin{lemma}[\sc Basic Matrix Multiplications]\label{lemma:matprod}
Let $A\in\mathbb{R}^{d\times q}$, $B\in\mathbb{R}^{q\times p}$, and $W\subseteq \{1,\ldots, d\}$. Then
\begin{enumerate}
    \item[(i).] $[A B]_{W.}=A_{W.}B$.
    \item[(ii).] If $d=q$, and $A$ is an invertible matrix, then $A_{W.}A^{-1}=I_{W.}$, where $I$ is the $d\times d$ identity matrix.
    \item[(iii).] If $d=q$, and $A$ is a diagonal matrix, then $A_{W.}B=A_{WW}B_{W.}$.
\end{enumerate}

\begin{proof}
    ~
    \begin{enumerate}
        \item[(i).] Simply note that $[A B]_{ij} = A_{i.} B_{.j}$, so 
        \[[A B]_{W.}=\{A_{i.} B_{.j}\}_{i\in W,j\in\{1,\ldots,q\}}=\{A_{i.} B\}_{i\in W}=A_{W.} B\]
        \item[(ii).] By (i), we have $A_{W.}A^{-1}=[AA^{-1}]_{W.}=I_{W.}$.
        \item[(iii).] Simply note that $A_{ik}=0$, for all $i\neq k$, so $A_{WW^c}=0$. Without loss of generality, suppose $W=1,\ldots,m$, with $1< m\le q$. Therefore, 
        \[A_{W.}B=[A_{WW},A_{WW^c}][B_{W.},B_{W^c.}]^\T =A_{WW}B_{W.}+A_{WW^c}B_{W^c.}=A_{WW}B_{W.} \] 
    \end{enumerate}
\end{proof}
\end{lemma}

\begin{lemma}\label{lemma:posdefSt}
For any $\Lambda\in\mathbb{R}^{d\times q}$ and $\Psi\in\mathcal{D}_{++}^{d\times d}$, with $d\ge q$, we have
    \[
    I_q-\Lambda^\T(\Lambda\Lambda^\T+\Psi)^{-1}\Lambda\succ 0
    \]
\begin{proof}
Let $R^\T$ be the matrix of eigenvectors of $\Lambda^\T\Psi^{-1}\Lambda$, and let $L = \Lambda R^\T$ be the rotated loading matrix. Thus, $D:=L^\T\Psi^{-1}L ={\rm diag}(a_1,\ldots,a_q)$ is a positive diagonal matrix by Lemma~\ref{lemma:canonrot}. Then
\begin{eqnarray*}
I_q-\Lambda^\T(\Lambda\Lambda^\T+\Psi)^{-1}\Lambda &=& R^\T R - R^\T L^\T(LRR^\T L^\T+\Psi)^{-1}LR\\
&=& R^\T(I_q-L^\T(LL^\T+\Psi)^{-1}L)R
\end{eqnarray*}
By expanding $(LL^\T+\Psi)^{-1}$ using the Woodbury Matrix Identity, we obtain
\begin{eqnarray*}
L^\T(LL^\T+\Psi)^{-1}L &=& L^\T\Psi^{-1}L-L^\T\Psi^{-1}L(I_q+L^\T\Psi^{-1}L)^{-1}L^\T\Psi^{-1}L\\
&=&D-D(I_q+D)^{-1}D\\
&=&{\rm diag}(a_1/(1+a_1),\ldots,a_q/(1+a_q))\\
&=&{\rm diag}(\alpha_1,\ldots,\alpha_q),
\end{eqnarray*}
where $0<\alpha_i<1$, for all $i=1,\ldots,q$. Therefore, $I_q-L^\T(LL^\T+\Psi)^{-1}L\succ 0$, and so is $I_q-\Lambda^\T(\Lambda\Lambda^\T+\Psi)^{-1}\Lambda$.
\end{proof}
\end{lemma}

~
\section{Start values for Algorithm~\ref{algo:mleEM}}
\label{app:startvalues}
We construct start values $(\Lambda_0,\Psi_0)$ for the LINFA MLE Algorithm~\ref{algo:mleEM} as follows. 
We first complete the observed data by filling in any missing data points about node $j$ with the simple average of all observed values about node $j$. This yields a $n\times d$ data matrix $\tilde{\mathbf{X}}$. We then obtain the sample covariance matrix $\tilde\Sigma$ and its spectral decomposition
\begin{equation}
\tilde\Sigma = V \Phi V^\T
\end{equation}
where $V\in\mathbb{R}^{d\times d}$ contains the eigenvectors, and $\Phi$ is the diagonal matrix containing the eigenvalues. The unrotated start value for $\Lambda$ is then defined as
\begin{equation}
    \tilde\Lambda_0 ~=~ V_{.,1:q}\Phi_{1:q,1:q}^{1/2}
\end{equation}
where $ V_{.,1:q}$ contains the first $q$ columns of $V$ and $\Phi_{1:q,1:q}^{1/2}={\rm diag}(\sqrt{\Phi_{11}},\ldots,\sqrt{\Phi_{qq}})$. The start value for $\Psi$ is defined as
\begin{equation}
\Psi_0 ~=~ {\rm Diag}(\tilde\Sigma)
\end{equation}
Finally, we compute the matrix $R$ of eigenvectors of $\frac{1}{d}\tilde\Lambda_0^\T\Psi_0^{-1}\tilde\Lambda_0$ to obtain the rotated start value
\begin{equation}
\Lambda_0 = \tilde\Lambda_0R
\end{equation}

\end{document}